\documentclass[twocolumn,superscriptaddress,longbibliography,aps,pra,preprintnumbers,10pt]{revtex4-1}
\usepackage{graphicx} 
\usepackage{bm}
\usepackage{color}
\usepackage{epstopdf}
\usepackage{amsmath}
\usepackage{amssymb}
\usepackage{ulem}
\usepackage{float}
\usepackage[urlcolor=blue,colorlinks=true,citecolor=blue,linkcolor=blue,pdfstartview={FitH},bookmarks=false]{hyperref}

\newcommand{\be}{\begin{equation}}
\newcommand{\ee}{\end{equation}}
\newcommand{\bea}{\begin{eqnarray}}
\newcommand{\eea}{\end{eqnarray}}

\newcommand{\w}{\omega}

\newcommand{\up}{\uparrow}
\newcommand{\down}{\downarrow}
\newcommand{\A}{\mathcal{A}}

\newcommand{\bk}{\mathbf{k}}

\newcommand{\nn}{\nonumber}
\newcommand{\mb}[1]{ { \mbox{\boldmath{$#1$}}}  } 

\newcommand{\tx}[1]{\text{#1}}

\begin{document}

\title{Influence of pairing and correlations on transport properties\\ in three-terminal superconducting nanostructure}

\author{G. G\'{o}rski}
\email{ggorski@ur.edu.pl}
\affiliation{Institute of Physics, Faculty of Exact and Technical Sciences, University of Rzesz\'{o}w, 35-310 Rzesz\'{o}w, Poland}

\author{V. Pokorn\'y}
\email{pokornyv@fzu.cz}
\affiliation{Institute of Physics (FZU), Czech Academy of Sciences, Na Slovance 2, CZ-18200 Praha 8, Czech Republic}

\author{T. Doma\'nski}
\email{doman@kft.umcs.lublin.pl}
\affiliation{Institute of Physics, M. Curie-Sk\l{}odowska University, 20-031 Lublin, Poland}

\date{\today}

\begin{abstract}
We study the local and nonlocal charge transport properties of three-terminal nanostructure, consisting of the correlated quantum dot embedded between two metallic electrodes and superconductor. Using the linear response theory, we analyze interplay of the induced on-dot pairing with the correlation effects, inspecting their influence on the conductance and thermopower. Our calculations predict negative values of the nonlocal conductance and reversal of the Seebeck coefficient when the quantum dot ground state is represented by the BCS-type singlet configuration. These phenomena are realized when the subgap charge transport is predominantly contributed by the crossed Andreev scattering. In contrast, on the doublet side, we obtain signatures for the Kondo effect manifested by the low-temperature enhancement of the charge conductance and clearly evident in the thermopower. Such temperature-dependent local and nonlocal transport properties can unambiguously identify the specific configuration of the correlated quantum dot and reveal the rich physics emerging in the doublet-to-singlet crossover region. 
\end{abstract}

\maketitle

\section{Introduction and motivation}
\label{sec.intro}

The transport properties of multiterminal superconducting hybrid structures with quantum dots (QDs) embedded in various interface geometries have recently become the subject of intensive theoretical and experimental research \cite{DeFranceschi-2014,Schindele-2014,Hashimoto-2024,Kouwenhoven-2025,Arrachea-2025}. Particular attention has been devoted to charge transport within the superconducting energy gap, where electron-to-hole (Andreev) scattering governs the low-energy tunneling processes \cite{Balatsky-2006,Rodero-2011}. In three-terminal junctions, these processes give rise to pronounced nonlocal effects mediated by crossed Andreev reflection (CAR). The interplay between single-electron tunneling (ET) and Andreev transport can result in negative nonlocal conductance \cite{Michalek-2013,Michalek-2016,Hashimoto-2024}, a sign reversal of the Seebeck coefficient \cite{Gorski-2025}, and a variety of other intriguing transport phenomena \cite{Governale-2026}. 

Quantum dots proximitized by bulk superconductors also provide a unique platform for exploring the competition between proximity-induced electron pairing and the local Coulomb repulsion. Depending on the coupling strength between the quantum dot and the superconductor and/or on the QD energy level, which can be tuned by gate voltages, the ground state may change between even- and odd-parity configurations \cite{Bauer-2007,Yamada-2011}. This quantum phase transition is accompanied by a crossing of the subgap bound states \cite{Balatsky-2006,Rodero-2011}, which has been observed experimentally using Josephson spectroscopy \cite{Maurand-2012,DeFranceschi-2014}, Andreev tunneling spectroscopy \cite{Jellinggaard-2016,Nygard-2020}, and numerous scanning tunneling microscopy (STM) measurements \cite{Heinrich-2018,Franke-2021}.

Furthermore, many-body correlations provide valuable insight into the interplay between localized electrons and the itinerant quasiparticles of the superconducting electrode(s). A prominent example is the Kondo effect, which influences both the differential conductance \cite{Deacon-2010,Franke-2011,Lee-2012,Pillet-2013} and the thermoelectric properties \cite{Arrachea-2025}. In particular, for a normal-metal--QD--superconductor junction, the subgap Kondo effect can enhance the zero-bias conductance as the quantum phase transition is approached from the doublet side \cite{Zitko-2015,Domanski-2016}. A similar enhancement has also been reported in Josephson junctions \cite{Deacon-2010,Lee-2012,Wu-2026}.

\begin{figure}
\centering
\includegraphics[width=0.95\linewidth]{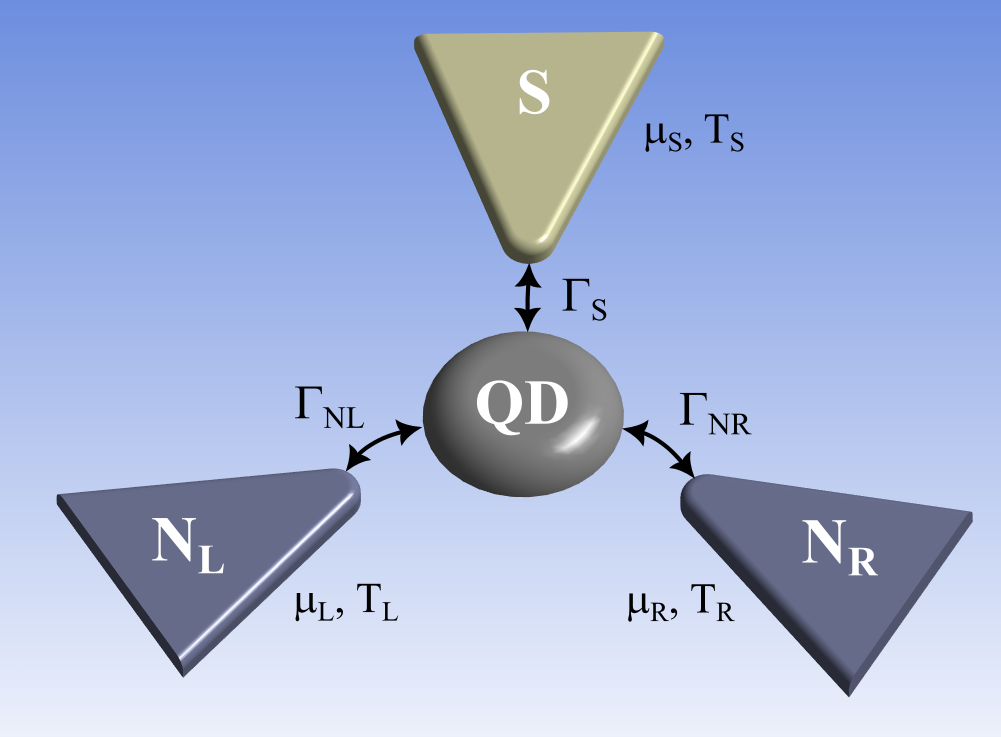}
\caption{Schematics of the three-terminal junction, consisting of the correlated quantum dot ($QD$) embedded between two metallic leads ($N_L$, $N_R$) and superconducting electrode ($S$).}
\label{scheme}
\end{figure}

The subtle interplay between proximity-induced on-dot pairing, Coulomb repulsion, and the subgap Kondo effect can be conveniently investigated in multiterminal hybrid junctions, where the various transport channels are experimentally well controlled. Of particular interest are three-terminal geometries in which the quantum dot is coupled either to two superconducting electrodes and one normal conducting lead \cite{Zhonda-2017} or to two normal metallic leads and one superconducting electrode \cite{Oguri-2026,Hashimoto-2024}. In this work, we focus on the latter configuration (Fig.~\ref{scheme}) and investigate the signatures of the parity transition and the subgap Kondo effect in the local and nonlocal conductance, as well as in the thermopower. 

This type of setup has been investigated using a variety of theoretical approaches \cite{Zhu-2001,Chen-2004,Das-2012,Michalek-2013,Wojcik-2014,Weymann-2014,Siqueira-2015,Michalek-2015,Michalek-2016,Gramich-2017,Trocha-2018,Pawlicki-2018,Belzig-2019,Flindt-2021,Hashimoto-2024,Sonar-2025}. It has been recognized as a promising platform for realizing Cooper-pair splitters and, more recently, as a minimal system capable of hosting Majorana-type quasiparticles \cite{Bulka-2026}. Influence of the subgap Kondo effect on the tunneling conductance has been recently investigated at zero temperature, $T = 0$, for the infinite pairing gap, $\Delta=\infty$, by means of  the numerical renormalization group calculations, Ref.~\cite{Hashimoto-2024}. Here we extend such studies to the finite gap,  $\Delta$, for various temperatures, and additionally investigate the thermoelectric properties.
We address this issue by analyzing the local and nonlocal transport properties, treating the Coulomb interaction within the self-consistent second-order perturbation theory. Thereby our study extends also the earlier results  \cite{Zhu-2001} obtained for arbitrary $\Delta$ in the junction with uncorrelated quantum dot,  by  demonstrating that the competition between ballistic electron transfer and anomalous Andreev transport channels allows to probe both the parity transition (crossover) and reveal hallmarks of the subgap Kondo effect. We identify their characteristic signatures appearing in the local/nonlocal conductance and thermopower and additionally discuss the role of magnetic field.

The paper is organized as follows. In Sec.~\ref{sec:model}, we introduce the microscopic model. The perturbative approach used to treat electron correlations is presented in Sec.~\ref{sec.perturbative}. In Sec.~\ref{sec:transport}, we define the transport coefficients within the linear-response framework. The numerical results obtained for the unpolarized junction are presented in Sec.~\ref{sec:results}, both at half-filling and away from particle--hole symmetry. In Sec.~\ref{sec:Exchange}, we discuss the effects induced by magnetic field for the half-filled quantum dot.  Sec.~\ref{sec:Conclusion} summarizes the main results and outlines possible directions for future research. The Appendices provide information on the influence of the superconducting gap (Appendix~\ref{gap_dependence}), a comparison with the unperturbed calculations (Appendix~\ref{sec:QMC}), and the effects of the magnetic field for the arbitrary energy level of QD (Appendix~\ref{Appendix_non_halfilled_QD}).

\section{Microscopic formulation}
\label{sec:formulation}

\subsection{Model Hamiltonian}
\label{sec:model}

The correlated QD coupled to two metallic leads ($N_L$, $N_R$) and superconducting ($S$) electrode can be described by the Anderson impurity model
\begin{eqnarray}
H=\sum_{\sigma}\varepsilon_{\sigma} d^{\dagger}_{\sigma} d_{\sigma} +U n_{d\uparrow}n_{d\downarrow}+ H_\tx{{N-QD}} +H_\tx{{S-QD}}, 
\label{HAM}
\end{eqnarray}
where $d_{\sigma}^\dagger$ ($d_{\sigma}$) are the creation (annihilation) operators of spin $\sigma$ electron at the energy level $\varepsilon_{\sigma}$,  $U$ denotes the on-dot repulsive Coulomb potential and $n_{d\sigma}=d^{\dagger}_{\sigma} d_{\sigma}$ is the electron number operator. The terms $H_{{\beta}-\tx{QD}}$ refer to the metallic $\beta=\tx{N}$ and superconducting $\beta=\tx{S}$ leads which are hybridized with QD. For the metallic electrodes they can be expressed as
\begin{equation}
H_\tx{{N-QD}} = \sum_{\mathbf{k},\sigma,\alpha} \xi_{\mathbf{k}{\alpha}}c^{\dagger}_{\mathbf{k}\sigma {\alpha}}c_{\mathbf{k}\sigma {\alpha}} + \sum_{\mathbf{k},\sigma,\alpha}(V_{\mathbf{k}{\alpha}}d^{\dagger}_{\sigma} c_{\mathbf{k}\sigma {\alpha}}+h.c.), 
\label{HNQD}
\end{equation}
where ${c}_{\mathbf{k}\sigma {\alpha}}^{\dag}$ $({c}_{\mathbf{k}\sigma {\alpha}})$
is the creation (annihilation) operator of the itinerant electron with spin $\sigma$, momentum $\mathbf{k}$ and  energy $\xi_{\mathbf{k}{\alpha}}=\varepsilon_{\mathbf{k}{\alpha}}-\mu_{\alpha}$ measured with respect to the chemical potential $\mu_{\alpha}$ of $\alpha=L,R$ metallic lead. In the wide bandwidth limit it is convenient to express the matrix elements $V_{\bk{\alpha}}$  in terms of  the coupling strength $\Gamma_{N{\alpha}}=2\pi\sum_{\bk}\left|V_{\bk {\alpha}}\right|^2\delta(\omega-\xi_{\bk{\alpha}})$ assumed to be $\omega$-independent in the energy regime of interest for our study.

The superconducting lead and its coupling to QD is described by
\bea
\label{HSQD}
H_\tx{{S-QD}} &=& \sum_{\bk,\sigma} \xi_{\bk S}c^{\dagger}_{\bk\sigma S}c_{\bk\sigma S} + \sum_{\bk,\sigma}(V_{\bk S}d^{\dagger}_{\sigma} c_{\bk\sigma S}+h.c.) \nonumber \\
&-&\sum_{\bk}(\Delta c^{\dagger}_{-\bk\uparrow S}c^{\dagger}_{\bk\downarrow S} + \Delta^{*}c_{\bk\downarrow S} c_{-\bk\uparrow S}) 
\eea
with the standard notation for $c_{\bk\sigma S}^{({\dagger})}$ operators, and $\Delta$ denotes the isotropic pairing gap. The hybridization of QD with the superconducting lead can be described by the coupling strength $\Gamma_{S}=2\pi\sum_{\bk,\sigma}\left|V_{\bk S}\right|^2\delta(\omega-\xi_{\bk S})$.   
This part (\ref{HSQD}) of the model Hamiltonian accounts for the on-dot electron pairing induced by the superconducting proximity effect that competes with the local Coulomb repulsion responsible for the correlation phenomena \cite{Bauer-2007}.

Transport properties of our three-terminal setup are determined by the quasiparticle features of the QD. To consider the induced on-dot pairing we introduce the Nambu spinor representation $\hat{\Psi}^\dag  = ( d_{\uparrow}^\dag, d_{\downarrow} )$, $\hat{\Psi}=(\hat{\Psi}^{\dagger})^{\dagger}$ and introduce the retarded/advanced Green's function
$\hat{\mb G}^{r/a}(\tau)=\left\langle \left\langle\hat{\Psi}(\tau) \left|\hat{\Psi}^\dag(0)\right.\right\rangle\right\rangle^{r/a}$.
Under equilibrium conditions its Fourier transform obeys the Dyson equation
\bea
\left[ \hat{\mb G}^{r/a}(\omega) \right]^{-1}=\left[\hat{\mb g}^{r/a}(\w)\right]^{-1}-\hat{\mb \Sigma}^{r/a}(\w)
\label{Dyson}
\eea
where $\hat{\mb g}^{r/a}(\w)$ refers to the retarded/advanced Green's function of the noninteracting case and $\hat{\mb \Sigma}^{r/a}(\w)$ is the matrix self-energy due to the Coulomb interaction. In absence of the correlations, $U=0$, the bare Green's functions $\hat{\mb g}^{r/a}(\w)$
can be obtained e.g.\ from the equation of motion method
\begin{eqnarray}
\omega \; \hat{\mb g}^{r/a}(\w)= \left\langle \left[\hat{\Psi},\hat{\Psi}^{\dagger}\right]_+ \right\rangle + \left\langle \left\langle \left[\hat{\Psi},H_{|_{U=0}}\right]_{-}\left| \hat{\Psi}^{\dagger}\right.\right\rangle\right\rangle^{r/a}_\omega
\label{EOM}
\end{eqnarray}
where $\left[A,B\right]_{\pm}$ stands for anticommutator/commutator. 
Exact solution of this equation (\ref{EOM}) for the noninteracting Hamiltonian (\ref{HAM}) gives \cite{Bauer-2007,Yamada-2011}
\begin{eqnarray}
\left[\hat{\mb g}^{r}(\w)\right]^{-1}=\left( \begin{array}{cc}
\w-\varepsilon_{\uparrow}+i\frac{\Gamma_{eff}}{2} &i\frac{\Gamma_S}{2}\beta(\w)\frac{\Delta}{\w}\\
i\frac{\Gamma_S}{2}\beta(\w)\frac{\Delta^*}{\w}&\w+\varepsilon_{\downarrow}+i\frac{\Gamma_{eff}}{2}
\end{array}\right),
\label{gr}
\end{eqnarray}
where
$\Gamma_{eff}=\Gamma_N+\Gamma_S\beta(\w)$, $\Gamma_N=\Gamma_{NL}+\Gamma_{NR}$, $\beta(\w)=\frac{\left|\w\right|}{\sqrt{\w^2-\Delta^2}}\theta\left(\left|\w\right|-\Delta\right)-i\frac{\w}{\sqrt{\Delta^2-\w^2}}\theta\left(\Delta-\left|\w\right|\right)$ and $\theta\left(x\right)$ is the Heaviside function. In the absence of the metallic leads, $\Gamma_N=0$, this Green's function (\ref{gr}) implies the resonant bound states existing inside the pairing gap. Their energies depend on the QD level $\varepsilon$ and the hybridization strength $\Gamma_{S}$. Specifically, in the strong coupling limit ($\Gamma_S \gg \Delta$) such bound states appear nearby the gap edges $\pm\Delta$, whereas for the weak hybridization ($\Gamma_S \ll \Delta$) they show up  inside the pairing gap at $E_{A\pm}=\pm \sqrt{\varepsilon^{2}+(\Gamma_S/2)^{2}}$. Presence of the metallic electrodes is responsible for a line-broadening of these in-gap states, with the inverse life-time $\tau\sim 1/\Gamma_{N}$ \cite{Baranski-2013}.

\subsection{Correlation effects}
\label{sec.perturbative}

The local Coulomb interactions $Un_{d\uparrow}n_{d\downarrow}$ compete with the proximity-induced on-dot pairing, $\langle d_{\downarrow}d_{\uparrow}\rangle$, and lead to the important correlation effects. Their role in various architectures of superconducting nanostructures has been investigated by numerous methods (for early reviews see Ref.\ \cite{Balatsky-2006,Rodero-2011}). To determine experimentally measurable transport properties in a broad range of the model parameters in three-terminal setup, we focus here on perturbative evaluation of the self-energy $\hat{\mb\Sigma}^{r/a}(\w)$ with respect to the Coulomb potential \cite{Yamada-2011,Zonda-2015,Domanski-2016}. Such formalism should be valid for the weakly correlated system, but it has proved to give fairly reliable results in considerations of the quantum phase transition and emergence of the subgap Kondo effect for two-terminal N-QD-S junctions \cite{Zonda-2015,Domanski-2016} and three-terminal structures as well \cite{Zhonda-2017}. In the next subsection we hence use it to compute the linear-response properties of our setup.

The diagonal and off-diagonal parts of the matrix self-energy $\hat{\mb\Sigma}^{r/a}(\w)$ are given by
\begin{eqnarray}
{\mb \Sigma}_{11}^{r/a}(\omega) &=&  U\langle
{d}^{\dagger}_{\downarrow}{d}_{\downarrow}\rangle 
 - \frac{U^2}{\pi} \int\limits_{-\infty}^{\infty}
{\frac{\mbox{\rm Im}{{\mb \Sigma}_{11}^{(2)}(\omega')}}
{{\omega-\omega'\pm i0^+}}d\omega'},
\label{diag_sigma} \\
{\mb \Sigma}_{12}^{r/a}(\omega)&=& U\langle
{d}_{\downarrow}{d}_{\uparrow}\rangle 
 + \frac{U^2}{\pi}\int\limits_{-\infty}^{\infty}
{\frac{\mbox{\rm Im}{{\mb \Sigma}_{12}^{(2)}(\omega')}}
{{\omega-\omega'\pm i0^+}}d\omega'} ,
\label{offdiag_sigma} 
\end{eqnarray}
with the second-order contributions
\begin{eqnarray}
\mbox{\rm Im}{\mb \Sigma}_{11(12)}^{(2)}(\omega)&=&-\int\limits_{-\infty}^{\infty}
\left[{\mb \Pi}_1(\omega+\omega')\rho_{22(21)}^+(\omega')\right.\nn\\
 &+&\left. {\mb \Pi}_2(\omega+\omega')\rho _{22(21)}^-(\omega')\right]d\omega' ,
\label{imag_part}
\\
{\mb \Pi}_{1(2)}(\omega)&=&\pi\int\limits_{- \infty}^{\infty}
\left[\rho_{11}^{-(+)}(\omega')\rho_{22}^{-(+)}(\omega-\omega')\right.\nn\\
 &-& \left.\rho_{12}^{-(+)}(\omega')\rho_{21}^{-(+)}(\omega-\omega') \right] d\omega' . 
\label{Pi_prop}
\end{eqnarray}
Here we use $\rho_{ij}^{\pm}(\omega)=-{\pi}^{-1} \mbox{\rm Im}{\mb G}^\tx{MF}_{ij}(\omega)f^{\pm}(\omega)$ where $f^{\pm}(\omega)=\left[ 1 + 
\mbox{\rm exp} \left(\pm \omega/(k_BT) \right)\right]^{-1}$ denotes 
the particle/hole Fermi-Dirac distribution function
and ${\mb G}^\tx{MF}_{ij}(\omega)$ is the Green's function obtained 
at the mean field approximation with static ($\omega$-independent) values of the diagonal $\Sigma_{11}^\tx{MF}$ and off-diagonal self-energy $\Sigma_{12}^\tx{MF}$ \cite{Yamada-2011}. The remaining parts of the matrix self-energy $\hat{\mb\Sigma}(\w)$ can be determined from the exact relations ${\mb \Sigma}_{22}(\omega )=-[{\mb \Sigma}_{11}(-\omega )]^{*}$ and 
${\mb \Sigma}_{21}(\omega )=[{\mb \Sigma}_{12}(-\omega )]^{*}$. 

\begin{figure}
\centering
\includegraphics[width=0.8\linewidth]{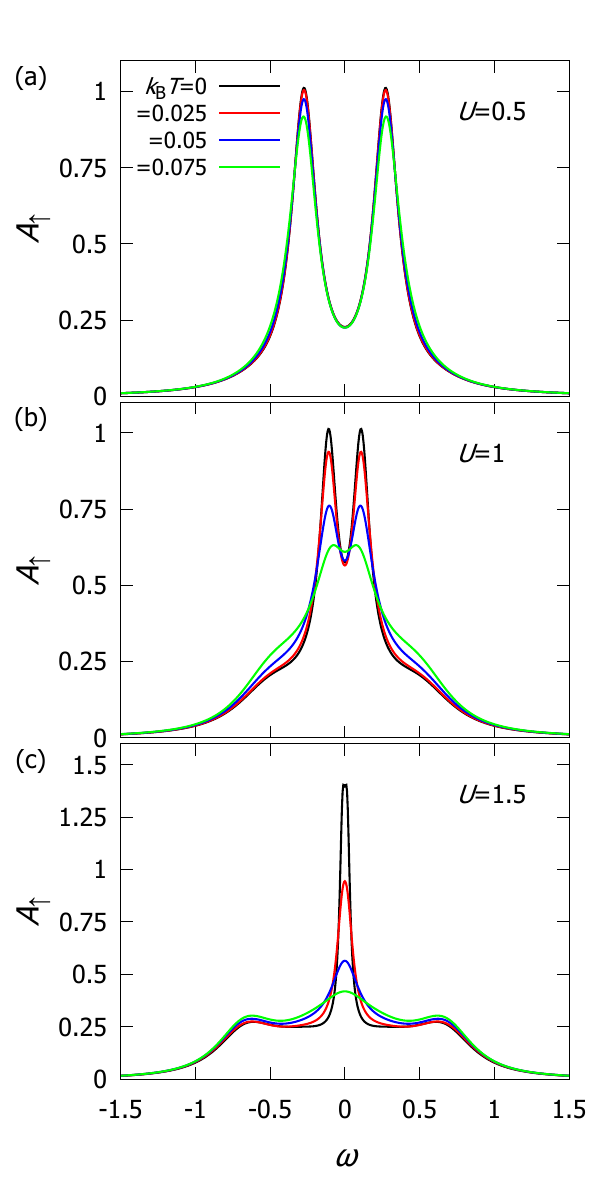}
\caption{
Normalized spectral function of the half-filled QD obtained for several temperatures, using the model parameters $\Gamma_{S}=1$, $\Gamma_{NL}=\Gamma_{NR}=0.125$, $\Delta=2$ and different values of $U$ as indicated.
}
\label{figSF}
\end{figure}

Interplay between the single particle and the Andreev transport channels (formulated in detail in Sec.\ \ref{sec:transport}) can be illustrated by the effective spectrum of the correlated QD. Fig.\ \ref{figSF} displays the normalized spectral function $\A_\up(\omega)=-\Gamma_N\mbox{\rm Im} G_{11}^{r}(\omega+i0^{+})$ obtained for three different values of the Coulomb potential $U$ and several temperatures. In the weak interaction limit ($U\leq \Gamma_S$) the ground state of QD is represented by the BCS-type configuration $u\left| 0\right> + v \left| \uparrow \downarrow \right>$. The spectrum consists of two peaks (broadened by $\Gamma_N$) at the energies of in-gap bound states, manifesting the particle and hole component of the Bogoliubov quasiparticles. In contrast, for the strong interaction limit ($U\geq \Gamma_S$) the ground state of QD is singly occupied $\left| \sigma \right>$, which is doubly degenerate in absence of magnetic field \cite{Bauer-2007}. Coupling of this singly occupied QD with itinerant electrons of the metallic electrodes along with the local Coulomb repulsion induces the effective spin-exchange interactions, which at low temperatures are responsible for the Kondo state \cite{Zitko-2015,Domanski-2016}. It is manifested in the spectral function by the Abrikosov-Suhl peak, appearing at $\omega=0$. Above the Kondo temperature, $T_K$, this peak disappears and QD spectrum evolves to two-peak structure.

To clarify the mechanism responsible for the subgap Kondo effect it is worth to recollect basic predictions of the Schrieffer-Wolff transformation formulated by us in Ref.\  \cite{Domanski-2016} for the large pairing gap limit, $\Delta \gg \Gamma_S$. This treatment implies that QD interacts with the mobile electrons of the normal $\alpha$-lead by the effective exchange potential
\begin{eqnarray}
J = \frac{|V_{k_{F}\alpha}|^{2}U}{\varepsilon (\varepsilon +U) +(\Gamma_S/2)^{2}}.
\end{eqnarray}
It yields the Kondo temperature
\begin{eqnarray}
T_{K} = \eta \frac{\sqrt{\Gamma_{N}U}}{2} \mbox{\rm exp} \left[ \frac{\varepsilon(\varepsilon+U)+(\Gamma_S/2)^{2}}{\Gamma_{N}U/\pi} \right] ,
\end{eqnarray}
where $\eta$ is dimensionless parameter of the order of unity.
Generalization of such canonical transformation to the finite pairing gap $\Delta$ has been discussed for two-terminal system by R. \v{Z}itko et al \cite{Zitko-2015}, and studied in great detail by means of the NRG calculations. 
Here, we check the validity of our perturbative treatment (\ref{diag_sigma},\ref{offdiag_sigma}), by presenting the results for the half-filled QD obtained by the unbiased quantum Monte Carlo (QMC) calculations [in Appendix \ref{sec:QMC}]. We find satisfactory agreement between both methods in the BCS region and in the regime of intermediate correlations ($U\leq\Gamma_S$). For $U\gg\Gamma_S$ the on-dot electron pairing is substantially reduced, and the Andreev mechanism is no longer contributing to the charge transport. 

In the next subsection we define the local and nonlocal transport coefficients. Numerical results are presented in Sec.\ \ref{sec:results}, where we discuss variation of the transport properties for a broad range of the model parameters, capturing the weak to intermediate interaction regions. We also emphasize signatures of the subgap Kondo effect.

\subsection{Charge transport}
\label{sec:transport}

\begin{figure}
\centering
\includegraphics[width=0.97\linewidth]{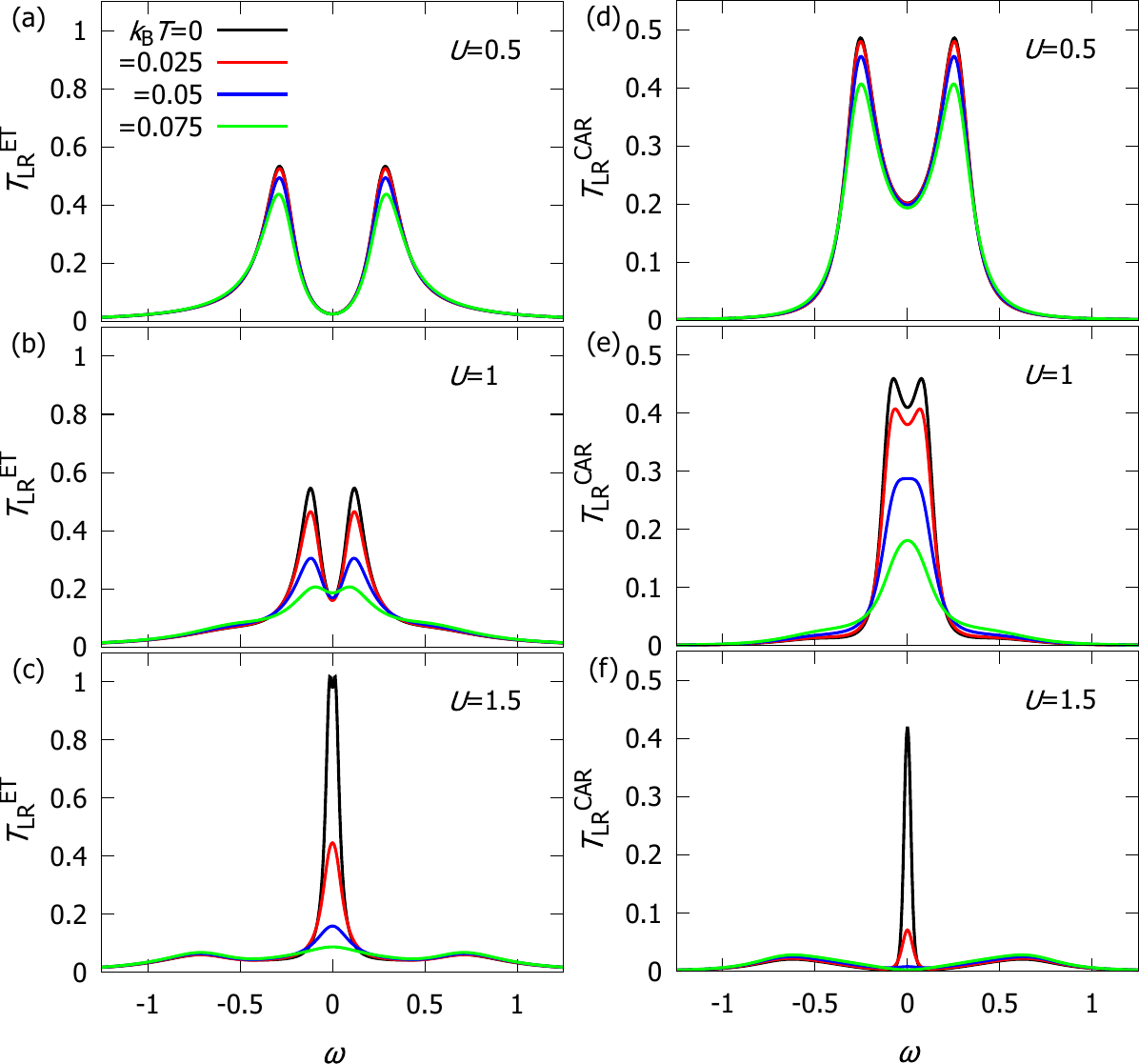}
\caption{Transmittance of the ET and CAR channels obtained for the half-filled QD, $\varepsilon=-\frac{U}{2}$, using the model parameters $\Gamma_{S}=1$, $\Gamma_{NL}=\Gamma_{NR}=0.125$ and $\Delta=2$.}
\label{diag_offdiag}
\end{figure}

The charge current flowing from the left normal lead is contributed by (i) the ballistic electron transfer (ET) from $L$ to $R$ lead, (ii) the Andreev reflection (AR) processes \cite{Michalek-2013,Michalek-2015,Michalek-2016}, and (iii) the quasiparticle current from $L$ to superconducting lead outside its pairing gap for high temperatures \cite{Hwang-2016,Verma-2022}. In our configuration with two metallic leads, there is a possibility of the direct Andreev reflection (DAR), when electron from $L$-th lead form a pair at the QD which propagates as the Cooper pair into superconductor S, back-scattering a hole of the opposite spin to the same lead. The other mechanism is the crossed Andreev reflection (CAR) which occurs when a hole is transferred to  $R$ lead \cite{Gorski-2025}. The spin-resolved total current can be expressed as
\bea
\label{Jstot}
J^\tx{TOT}_{L\sigma}= J^\tx{ET}_{L\sigma}+J^\tx{DAR}_{L\sigma} +J^\tx{CAR}_{L\sigma}+J^\tx{QP}_{L\sigma},
\eea
where
\bea
\label{JsET}
J^\tx{ET}_{L\sigma} &=& \frac{e}{h}\int{d\omega T^\tx{ET}_{LR\sigma}(\omega)\left[f_{L}(\omega)-f_{R}(\omega)\right]},
\\
\label{JsDAR}
J^\tx{DAR}_{L\sigma}&=& \frac{e}{h}\int{d\omega T^\tx{DAR}_{LL\sigma}(\omega)\left[f_{L}(\omega)-\bar{f}_{L}(\omega)\right]},
\\
\label{JsCAR}
J^\tx{CAR}_{L\sigma} &=& \frac{e}{h}\int{d\omega T^\tx{CAR}_{LR\sigma}(\omega)\left[f_{L}(\omega)-\bar{f}_{R}(\omega)\right]},
\\
\label{JsQP}
J^\tx{QP}_{L\sigma}&=& \frac{e}{h}\int{d\omega T^\tx{QP}_{LS\sigma}(\omega)\left[f_{L}(\omega)-f_{S}(\omega)\right]},
\eea
where $f_{\alpha}(\omega)=\left\{[1+{\rm exp}\left[(\omega-\mu_\alpha)/(k_BT_\alpha)\right]\right\}^{-1}$  denotes electron and 
$\bar{f}_{\alpha}(\omega)=\left\{[1+{\rm exp}\left[(\omega+\mu_\alpha)/(k_BT_\alpha)\right]\right\}^{-1}$ hole distribution function, respectively.
The transmission coefficients $T^\kappa_{\alpha\beta\sigma}$ ($\kappa=\left\{\text{ET},\text{DAR},\text{CAR}\right\}$) are defined by
\bea
\label{TET}
T^\tx{ET}_{LR\sigma}(\omega)&=&\Gamma_{NL}\Gamma_{NR}\left|\left\langle \left\langle d_{\sigma}\left|d_{\sigma}^\dag\right.\right\rangle\right\rangle^r_\omega\right|^2, 
\\
\label{TDAR}
T^\tx{DAR}_{LL\sigma}(\omega)&=&\Gamma_{NL}\Gamma_{NL}\left|\left\langle \left\langle d_{\sigma}\left|d_{-\sigma}\right.\right\rangle\right\rangle^r_\omega\right|^2, 
\\
\label{TCAR}
T^\tx{CAR}_{LR\sigma}(\omega)&=&\Gamma_{NL}\Gamma_{NR}\left|\left\langle \left\langle d_{\sigma}\left|d_{-\sigma}\right.\right\rangle\right\rangle^r_\omega\right|^2,
\\
\label{TQP}
T^\tx{QP}_{LS\sigma}(\omega)&=&\Gamma_{NL}\Gamma_{S}\frac{{\left|\w\right|\theta\left(\left|\w\right|-\Delta\right)}}{\sqrt{\w^2-\Delta^2}}\left\{\left|\left\langle \left\langle d_{\sigma}\left|d^{\dagger}_{\sigma}\right.\right\rangle\right\rangle^r_\omega\right|^2 \right.\nn \\
 &+& \left|\left\langle \left\langle d_{\sigma}\left|d_{-\sigma}\right.\right\rangle\right\rangle^r_\omega\right|^2
\\ &-&\left.\frac{2\Delta}{\left|\w\right|}\text{Re}\left[\left\langle \left\langle d_{\sigma}\left|d^{\dagger}_{\sigma}\right.\right\rangle\right\rangle^r_\omega \left[\left\langle \left\langle d_{\sigma}\left|d_{-\sigma}\right.\right\rangle\right\rangle^r_\omega\right]^*\right]\right\}\nn.
\eea 
In similar manner we can express the current $J_R$, exchanging the indices $L\leftrightarrow R$. In Fig.\ \ref{diag_offdiag} we show the transmittance of CAR and ET channels for the half-filled QD. 

In the linear response regime, the electrodes are detuned by infinitesimal perturbations $\mu_\alpha=\mu_S+\delta \mu_\alpha$ and $T_\alpha=T_S+\delta T_\alpha=T+\delta T_\alpha$. Under such conditions, the charge transport from $L$-lead simplifies to \cite{Michalek-2016}
 \bea
\label{JsLtotdel}
J_{L\sigma}&=& D_{LR,\mu_\sigma}^\tx{ET}(\delta \mu_L-\delta \mu_R)+
2D_{LL,\mu_\sigma}^\tx{DAR}\delta \mu_L \\
&+&D_{LR,\mu_\sigma}^\tx{CAR}(\delta \mu_L+\delta \mu_R)+D_{LS,\mu_\sigma}^\tx{QP}\delta \mu_L\nn \\
&+&D_{LS,T}^\tx{QP}\delta T_L+(D_{LR,T}^\tx{ET}+D_{LR,T}^\tx{CAR})(\delta T_L-\delta T_R),\nn
\eea
where the coefficients $D_{\alpha\beta,\mu_\sigma}^{\kappa}$ and $D_{\alpha\beta,T}^{\kappa}$ have the following meaning
\bea
\label{DLRmu}
D_{\alpha\beta,\mu_\sigma}^{\kappa}&=&\frac{e}{\hbar}\int {d\omega T^{\kappa}_{\alpha\beta\sigma}(\omega)\left[-\displaystyle \frac{\partial f}{\partial \omega}\right]},
\\
\label{DLRT}
D_{\alpha\beta,T}^{\kappa}&=&\frac{e}{ \hbar T}\int {d\omega \omega T^{\kappa}_{\alpha\beta\sigma}(\omega)\left[-\displaystyle \frac{\partial f}{\partial \omega}\right]}.
\eea
The chemical potentials can be practically controlled by external gate-potentials, $\delta 
 \mu_\alpha=e V_\alpha$. In what follows, we assume the superconducting lead to be grounded, $\mu_S=0$.

In multi-terminal junctions the transport coefficients (the linear conductance and Seebeck coefficient)  are sensitive to the way in which the bias $V_{\alpha}$ or the temperature difference is imposed. For our system we can define the following transport coefficients \cite{Mazza-2014,Michalek-2013,Michalek-2016}: the electrical conductance $G_{LL\sigma}=\lim_{\delta V_L\rightarrow 0} \left.{\frac{J_{L\sigma}}{\delta V_L}}\right|_{\delta T=0} $ (local) and $G_{RL\sigma}= -\lim_{\delta V_L\rightarrow 0}\left.{\frac{J_{R\sigma}}{\delta V_L}}\right|_{\delta T=0} $ (nonlocal) and the thermoelectric power $ S_{LL\sigma}= -\lim_{\delta T_L\rightarrow 0}\left.{\frac{\delta V_L}{\delta T_L}}\right|_{J_{\alpha\sigma}=0} $ (local) and $ S_{RL\sigma}=- \lim_{\delta T_L\rightarrow 0}\left.{\frac{\delta V_R}{\delta T_L}}\right|_{J_{\alpha\sigma}=0} $ (nonlocal), respectively.

Using Eq.\ (\ref{JsLtotdel}) we obtain
\bea
\label{GLLsigma}
G_{LL\sigma}=e\left[D_{LR,\mu_\sigma}^\tx{ET}+2D_{LL,\mu_\sigma}^\tx{DAR}+D_{LR,\mu_\sigma}^\tx{CAR}+D_{LS,\mu_\sigma}^\tx{QP}\right],
\eea
and
\bea
\label{GRLsigma}
G_{RL\sigma}=e\left[D_{LR,\mu_\sigma}^\tx{ET} - D_{LR,\mu_\sigma}^\tx{CAR}\right].
\eea
In presence of an external magnetic field ($B \neq 0$) the differential conductance becomes spin-dependent, $G_{\alpha \beta\up} \neq G_{\alpha \beta\down}$, where $\alpha,\beta$ stand for $L$ and $R$. Under such conditions it is useful to define the total electrical conductance
\bea
\label{GLLGRLTOT}
G_{\alpha \beta}=G_{\alpha \beta\up} + G_{\alpha \beta\down},
\eea
and its spin-polarization
\bea
\label{PGLLGRL}
G^S_{\alpha \beta}=G_{\alpha \beta\up} - G_{\alpha \beta\down}.
\eea

The local and nonlocal Seebeck coefficients of our three-terminal setup are expressed in the linear response regime by
\bea
\label{SLLsigma}
S_{LL\sigma}&=&\frac{D_{\sigma T_L}^{ET}+D_{\sigma T_L}^{CAR}}{G_{LL\sigma}},
\\
\label{SRLsigma}
S_{RL\sigma} (T)&=&-\frac{D_{\sigma T_L}^{ET}+D_{\sigma T_L}^{CAR}}{G_{RL\sigma}}.
\eea
To capture their spin-dependence, we introduce the averaged thermopower 
\bea
\label{eq:S}
{S_{\alpha \beta}}=\frac{S_{\alpha \beta\up} + S_{\alpha \beta\down}}{2}
\eea
and its spin-polarized component
\bea
\label{eq:SS}
{S^S_{\alpha \beta}}=\frac{S_{\alpha \beta\up} - S_{\alpha \beta\down}}{2}
\eea
adopting  the convention used in Refs.\  \cite{Swirkowicz-2009,Weymann-2013}.

At low temperature the current contributed by quasiparticles from outside the pairing gap of the superconducting lead would be negligible ($D_{LS,\mu_\sigma}^\tx{QP}=0$), therefore $G_{LL}$ and $G_{RL}$ will arise from the ET and the AR processes. For $T=0$, the local and nonlocal linear conductance is given by 
\bea
\label{GLL0}
G_{LL}=8\frac{e^2}{\hbar}z^2\frac{\Gamma_{NL}\Gamma_{NR}(4\tilde{\varepsilon}^2+\tilde{\Gamma}_{N}^2+\tilde{\Gamma}_{S}^2)+2\Gamma_{NL}^2\tilde{\Gamma}_{S}^2}{(4\tilde{E}_A^2+\tilde{\Gamma}_{N}^2)^2},
\eea
and
\bea
\label{GRL0}
G_{RL}=8\frac{e^2}{\hbar}z^2\frac{\Gamma_{NL}\Gamma_{NR}(4\tilde{\varepsilon}^2+\tilde{\Gamma}_{N}^2-\tilde{\Gamma}_{S}^2)}{(4\tilde{E}_A^2+\tilde{\Gamma}_{N}^2)^2},
\eea
where $\tilde{\Gamma}_{N}$ and $\tilde{\Gamma}_{S}$ are  defined by \cite{Tanaka-2007, Yamada-2011} 

\bea
\label{GammaNeff}
\tilde{\Gamma}_{N}&=&z{\Gamma}_{N},
\\
\label{GammaSeff}
\tilde{\Gamma}_{S}&=&z\left[{\Gamma}_{S}-2\Sigma_{12}(0)\right],
\eea
with the renormalization factor 
\bea
\label{ZZ}
z=\left(1+\frac{{\Gamma}_{S}}{2\Delta}-\left.\frac{\partial \Sigma_{11}(\omega)}{\partial \omega}\right|_{\omega=0}\right)^{-1}.
\eea
The renormalized QD energy $\tilde{\varepsilon}$ is
\bea
\label{epsilonDeff}
\tilde{\varepsilon}=z\left[{\varepsilon}+\Sigma_{11}(0)\right],
\eea
implying the in-gap bound states at
\bea
\label{EDeff}
\tilde{E}_{A\pm}=\pm \sqrt{\tilde{\varepsilon}^2+\tilde{\Gamma}_{S}^2/4}.
\eea
These features are evident in the spectral function (Fig.~\ref{figSF}) and indirectly also in the transmittance
(Fig.~\ref{diag_offdiag}).

\section{Results for unpolarized setup}
\label{sec:results}

\begin{figure}[b!]
\centering
\includegraphics[width=0.9\linewidth]{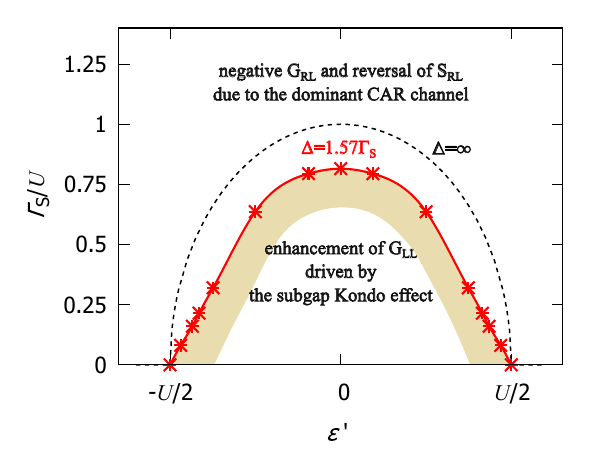}
\caption{
Diagram obtained for $\Gamma_{N}=0$, showing the singlet-double phase boundary for $\Delta=\infty$ (black dashed line) and for finite pairing gap (red line with points) Ref.~\cite{Bauer-2007}. In the case $\Gamma_{N}\neq 0$ relevant to our setup the quantum phase transition evolves into crossover (marked in color). We indicate the regions where the negative nonlocal conductance and the reversed Seebeck coefficients occur due to the dominant CAR transport channel which occur on the BCS-singlet side, and the subgap Kondo effect which strongly enhances the local conductance in the crossover region \cite{Zitko-2015,Domanski-2016}.}
\label{diagram}
\end{figure}

We begin by presenting the numerical results for the transport coefficients obtained in the absence of an external magnetic field. For convenience, we treat the QD coupling to superconductor as a unit for the energies, $\Gamma_{S}\equiv 1$ and focus on the weak hybridization of QD with the metallic leads, $\Gamma_{N}< \Delta$, which guarantees a small broadening of the in-gap bound states, yet strong enough for emergence of the subgap Kondo state.

We determined the transport properties within perturbative treatment of the Coulomb potential. Such results should be valid in the weak and intermediate interaction regime for arbitrary model parameters ($\varepsilon$, $\Delta$, $\Gamma_{N}$, $\Gamma_{S}$, and temperature). The main findings are displayed in Fig.~\ref{diagram}.

\subsection{Conductance of half-filled quantum dot}
\label{half_filled_QD}

Let us start by discussing the signatures of the Kondo effect observable in transport properties for the particle-hole symmetric case, $\varepsilon=-\frac{U}{2}$. In such situation, the thermoelectric power vanishes, we hence restrict our considerations to the local $G_{LL}$ and nonlocal $G_{RL}$ conductance. Similar results have been recently considered  in Refs.\ \cite{Hashimoto-2024,Oguri-2026} for $\Delta\rightarrow\infty$. Here we analyze variation of the transport properties against the Coulomb potential $U$ for arbitrary $\Delta/\Gamma_S$ ratio, indicating hallmarks of the subgap Kondo effect.

\begin{figure}
\centering
\includegraphics[width=0.95\linewidth]{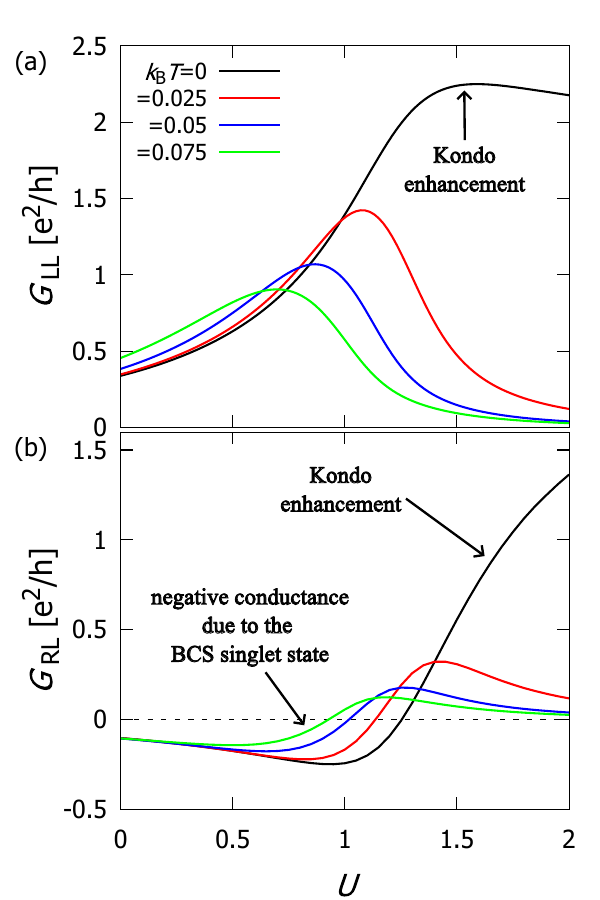}
\caption{
Variation of the (a) local $G_{LL}$  and (b) nonlocal $G_{RL}$ linear conductance with respect to the interaction potential $U$ obtained for several  temperatures (indicated in legend). Results are obtained for the half-filled QD, $\varepsilon=-\frac{U}{2}$, using the model parameters $\Gamma_{S}=1$, $\Gamma_{NL}=\Gamma_{NR}=0.125$ and $\Delta=2$.
}
\label{figGUTwnd1}
\end{figure}

Figure \ref{figGUTwnd1}(a) shows variation of the local conductance versus $U$ for several temperatures, as indicated. At temperatures well below the critical $T_c=\Delta/1.76$ the tunneling conductance comes mainly from the ET and AR transport channels, because the quasiparticles from outside the pairing gap have a rather negligible efficiency. The relative ET/AR contributions are strongly dependent on the Coulomb potential, which determines character of the in-gap bound states. In the weak coupling regime, $U\leq \Gamma_S$, the QD is in the BCS-singlet configuration, therefore the Andreev mechanism plays a dominant role. In contrast, for $U \geq \Gamma_S$, the doublet configuration of QD promotes the single-particle transport. In the latter case, by approaching such phase boundary (which is smeared into {\it crossover} due to $\Gamma_N\neq 0$) this ET channel conductance can be substantially enhanced by the Kondo state \cite{Zitko-2015,Domanski-2016,Yamada-2011}. Efficiency of this Kondo enhancement is sensitive to temperature (Fig.\ \ref{figSF}). 
Deep in the Kondo regime, the optimal conductance $G_{LL}$ exceeds $2e^{2}/h$ [black curve in Fig.\ \ref{figGUTwnd1}(a)],  due to the enhanced single-electron (ET) transport combined with additional contribution from the direct Andreev reflection (DAR) channel. With increase of temperature the Kondo state disappears, thereby the local conductance substantially diminishes above the singlet-doublet crossover, $U>\Gamma_S$. This behavior can be understood by inspecting the transmittance displayed in Fig.~\ref{diag_offdiag}.

Figure \ref{figGUTwnd1}(b) shows the nonlocal conductance $G_{RL}$ varying with respect to the Coulomb interaction $U$ for the same set of temperatures as in panel a. The current flowing into the right metallic lead is sensitive to the voltage applied to the left normal electrode mainly due to the crossed Andreev reflections. In particular, the nonlocal conductance $G_{RL}$ acquires negative values in the region corresponding to the BCS-singlet configuration \cite{Michalek-2015,Hashimoto-2024}. Let us remark that the Kondo effect is then absent. In contrast, on the doublet side (for $U\geq\Gamma_S$) the dominant contribution to nonlocal charge transport comes from the ET channel; therefore, the conductance changes sign to positive values. In such a limit of the strong Coulomb interactions ($U>\Gamma_S$) the subgap Kondo physics is again relevant \cite{Domanski-2016}, enhancing the nonlocal conductance at low temperatures. 

The nonlocal conductance is thus particularly valuable tool capable to  detect the singlet-doublet changeover (featured by negative values of $G_{RL}$ on the BCS-singlet side) and presence of the subgap Kondo state (manifested by temperature sensitivity of the positive-valued conductance on the doublet side). Our conclusions concerning the BCS-singlet region are perfectly consistent with the nonperturbative results obtained recently for the same three-terminal setup by M. Hashimoto et al \cite{Hashimoto-2024} by means of NRG calculations, assuming the infinite pairing gap $\Delta\rightarrow\infty$. Our present study generalizes their results on the case of arbitrary pairing gap (Appendix \ref{gap_dependence} shows variation on the local and nonlocal conductance obtained for various $\Delta$) and emphasizes additional features of the subgap Kondo effect appearing on the doublet side. These phenomena could be verified experimentally.

\subsection{Transport properties away from half-filling}
\label{sec:IIIB}

We now analyze variation of the transport properties outside the particle-hole symmetric case. To this end we introduce the {\it detuning} parameter $\varepsilon'=\varepsilon+\frac{U}{2}$. In practice, the energy level of QD can be experimentally lifted/lowered from $\varepsilon'=0$ by applying the external gate-potential. As we shall see, departure from the particle-hole symmetry leads to qualitative changes of the conductance and thermopower. Specifically, for the strongly correlated QD a changeover from the Kondo to the mixed valence regime would occur. Further important effects would be related to the proximity induced on-dot electron pairing. We discuss these phenomena for a few sets of $U/\Gamma_S$ ratios and various temperatures.

\begin{figure}
\centering
\includegraphics[width=0.95\linewidth]{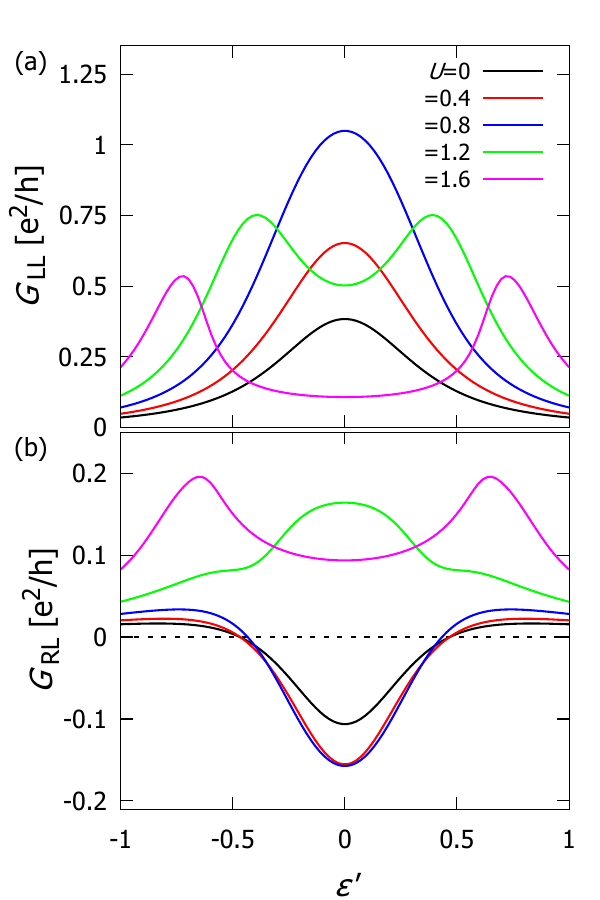}
\caption{
 (a) The local $G_{LL}$  and (b) nonlocal $G_{RL}$ linear conductance as a function of the detuning parameter  $\varepsilon'$ for several values of the Coulomb interaction $U$ obtained for $k_BT=0.05$, $\Gamma_{S}=1$, $\Gamma_{NL}=\Gamma_{NR}=0.125$ and $\Delta=2$.
}
\label{figGedprimkU}
\end{figure}

Let us start by discussing the local and nonlocal conductance in the region outside the particle-hole symmetry case, $\varepsilon'\neq 0$. In the weak interaction limit,  $U<\Gamma_S$, the induced on-dot pairing implies the local conductance (Fig.\ \ref{figGedprimkU}a) characterized by a single-peak centered at $\varepsilon'=0$. In contrast, in the strong interaction limit, this conductance acquires two-peak structure positioned roughly at $\varepsilon'=\pm U/2$. Such qualitative evolution is determined by the low-energy transmittance which in the weak interaction limit is contributed mainly by the Andreev reflections, and otherwise by the ET channel.

Additional evidence for this changeover is seen in the nonlocal conductance (Fig.\ \ref{figGedprimkU}b). In Sec.\ \ref{half_filled_QD} we have shown that $G_{RL}$ acquires negative values when the crossed Andreev scattering dominates over the single electron tunneling. In general, such situation occurs for such $\varepsilon`$ which coincide with the BCS-type ground state of QD \cite{Bauer-2007}. In agreement with previous results obtained for $\Delta\rightarrow\infty$ in Ref.\ \cite{Hashimoto-2024} we find the most remarkable negative values of $G_{RL}$ nearby the half-filling. In the strongly interacting limit, the nonlocal conductance changes sign to positive values because the charge current is mainly contributed by the ET channel. Temperature is also relevant for the latter case due to the subgap Kondo effect. 

\begin{figure}
\centering
\includegraphics[width=0.95\linewidth]{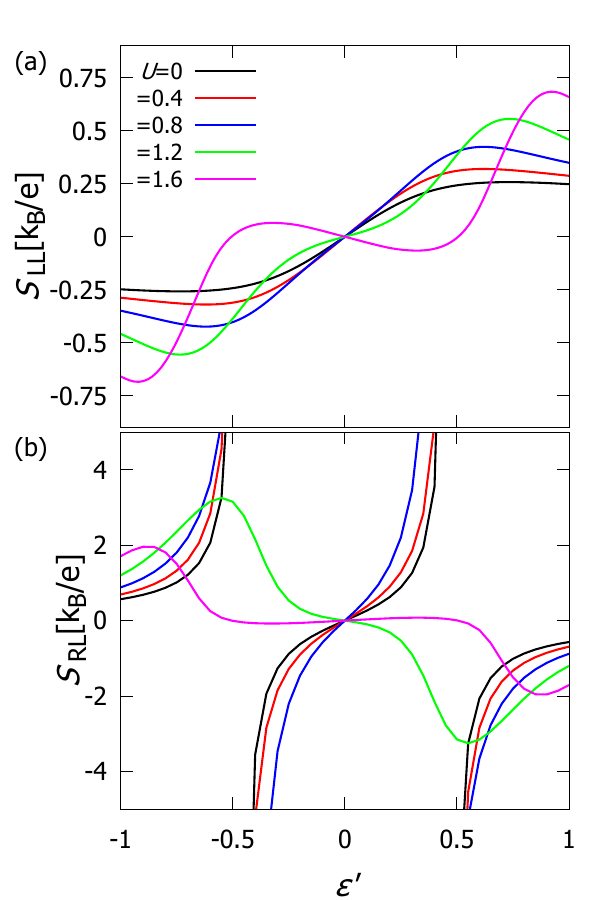}
\caption{
Variation of (a) the local $S_{LL}$ and (b) nonlocal $S_{RL}$ Seebeck coefficient with respect to $\varepsilon'=\varepsilon+\frac{U}{2}$ obtained for the same set of model parameters as in Fig.\ \ref{figGedprimkU}.
}
\label{figSedprimkU}
\end{figure}

Complementary signatures of these changeover effects show up in the thermopower. The local Seebeck coefficient (Fig.\ \ref{figSedprimkU}a) has {\it saw-tooth} shape as a function of the QD energy $\varepsilon'$, where its sign indicates either the particle or hole dominated transport \cite{Michalek-2015,Michalek-2016}. However, we notice that such a shape is reversed when entering the strong interaction limit, $U > \Gamma_S$. 
This puzzling reversal is also accompanied with qualitative change of the nonlocal thermopower, Fig.\ \ref{figSedprimkU}b. In the weak interaction limit, $U<\Gamma_{S}$, the nonlocal thermopower has two divergence points. Divergence of the nonlocal thermopower coincides with the sign-change of the nonlocal conductance (Fig.\ \ref{figGedprimkU}b) and both of them are hallmarks of the ground state parity change. In the strong interaction limit, $U\gg \Gamma_{S}$, such divergence points no longer exist. In the latter case, $S_{LR}$ changes its sign only once, exactly at the half-filling  in accordance with the Mott relation \cite{Mahan-1980}. For $U\sim\Gamma_S$ we observe a smooth evolution between these qualitatively different shapes. Sign-reversal of the derivative $\partial S_{RL}/\partial \varepsilon'$ can be hence considered as very sensitive tool for identifying the type of QD ground state.

\begin{figure} 
\centering
\includegraphics[width=0.95\linewidth]{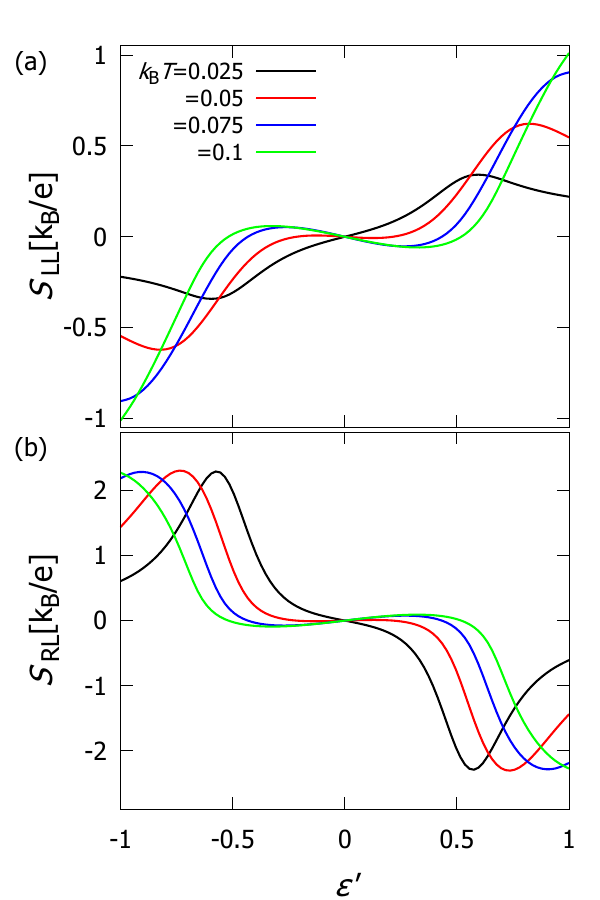}
\caption{
(a) The local $S_{LL}$ and (b) nonlocal $S_{RL}$ thermoelectric power as a function of  $\varepsilon'=\varepsilon+\frac{U}{2}$ obtained for $U=1.4$, $\Gamma_{NL}=\Gamma_{NR}=0.125$ and $\Delta=2$ and several temperatures indicated in the legend [in units of $\Gamma_{S}$].
}
\label{figSedprimkTS}
\end{figure}

Further modifications arise close to the half-filling, when the subgap Kondo effect is realized at low temperatures in the strong interaction limit. They are manifested by temperature-dependence of the Seebeck coefficients. Fig.\ \ref{figSedprimkTS} displays the local (panel a) and nonlocal (panel b) thermopower for several temperatures. 

\begin{figure*}[th]
\centering
\includegraphics[width=1\linewidth]{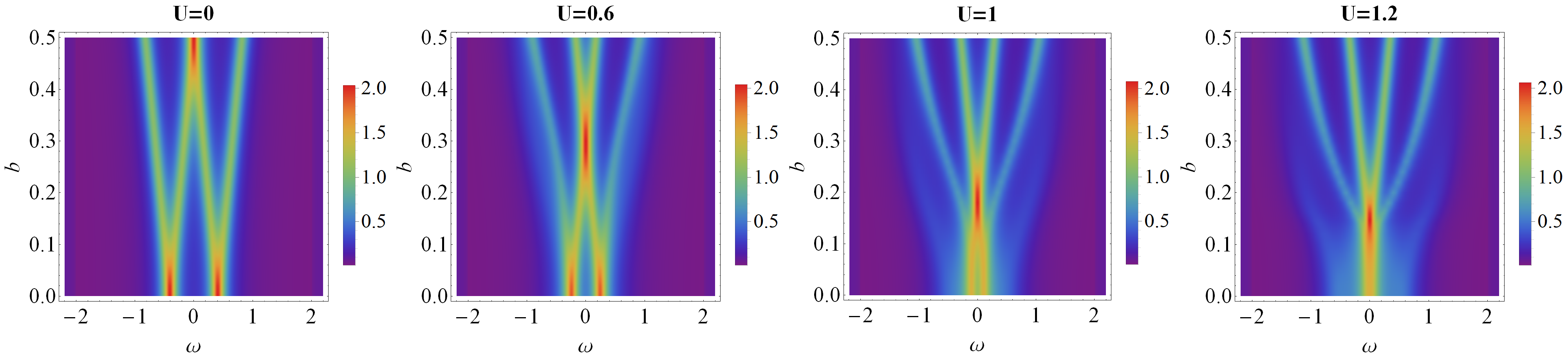}
\caption{
Variation of the spectral function of the half-filled QD with respect to the Zeeman energy, $b$, obtained for several values of the Coulomb interaction (as indicated), using the model parameters $\Gamma_{S}=1$, $\Gamma_{NL}=\Gamma_{NR}=0.125$, $\Delta=2$, and assuming $k_BT=0.05$.
}
\label{figAkB}
\end{figure*}

We see that the Kondo effect has qualitative influence of the local Seebeck coefficient (panel a), leading to sign-reversal of $S_{LL}$ at temperatures below $T_{K}(\varepsilon')$. As concerns the nonlocal thermopower (panel b) it merely flattens the small-valued $S_{LR}$. The latter effect is rather obvious, if we recall that nonlocal transport phenomena originate from the crossed Andreev reflection and these are strongly suppressed in the doublet region (where the subgap Kondo effect can appear). For this reason the nonlocal Seebeck coefficient is rather weakly sensitive to presence of the Kondo peak at low temperatures.

\section{Magnetic field effects}
\label{sec:Exchange}

We finally address how an external magnetic field ($B$) affects the local and nonlocal transport coefficients of our three-terminal superconducting setup. Its influence on the quantum dot lifts the spin degeneracy of the quasiparticle states through the Zeeman splitting, $(\varepsilon_\downarrow-\varepsilon_{\uparrow})/2= b\equiv \mu_B B$, where $\mu_B$ denotes the Bohr magneton. Depending on a magnitude of $b$, the linear conductance and Seebeck coefficient are modified in both the weak-correlation regime, where transport is dominated by Andreev processes, and the strong-correlation regime, where the single-particle transport channel prevails.

For brevity, we focus here on the half-filled QD case, $\frac{1}{2}\left(\varepsilon_{\uparrow}+\varepsilon_{\downarrow}\right)=-\frac{U}{2}$. Role of the magnetic field for arbitrary filling of QD is discussed in Appendix \ref{Appendix_non_halfilled_QD}. We begin by inspecting variation of the spectral function with respect to the Zeeman energy $b$ for selected values of the Coulomb potential (Fig.~\ref{figAkB}). The weak interaction spectrum consists of two spin-degenerate Andreev states, $E_{A\pm}$, which via the Zeeman splitting evolve into $E^\up_{A\pm}$ and $E^\down_{A\pm}$, respectively. At some critical value, $b_{crit}$, a pair of the internal quasiparticle states ($E^\up_{A-}$ and $E^\down_{A+}$) cross each other, enhancing  the zero-energy spectral function, $A(\omega=0)$. In contrast to this behavior, in the strong interaction regime the QD spectral function is represented by the central peak (due to the subgap Kondo state) coexisting with the  Coulomb satellites (Fig. \ref{figSF}). In  presence of the weak magnetic field there occurs a transfer of the spectral weights between such quasiparticle states. Upon further increasing $b$ the Kondo state ultimately disappears and we observe gradual reconstruction of the four-peak structure. The latter effect shows that magnetic field shifts the singlet-doublet transition, in accordance to what has been pointed out previously in Ref.~\cite{Hashimoto-2024}.

\begin{figure}[b!]
\centering
\includegraphics[width=0.95\linewidth]{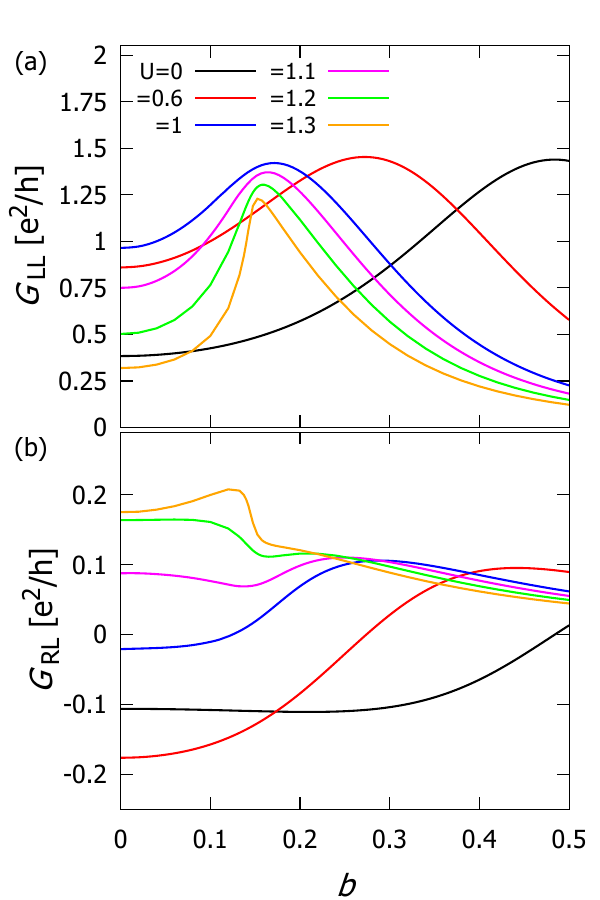}
\caption{
(a) The local $G_{LL}$ and (b) the nonlocal conductance $G_{RL}$ as functions of the Zeeman energy $b$ for several values of Coulomb interaction $U$., using the model parameters $\Gamma_{S}=1$, $\Gamma_{NL}=\Gamma_{NR}=0.125$, $\Delta=2$, $\varepsilon=-\frac{U}{2}$ and $k_BT=0.05$.}
\label{figGLLRLkBB}
\end{figure}

Qualitative variation of the low-energy QD spectrum induced by the magnetic field indirectly affects the transport properties. Fig.~\ref{figGLLRLkBB} shows the local ($G_{LL}$) and nonlocal ($G_{RL}$) conductance, varying against the Zeeman energy for several values of the Coulomb interaction, $U$. In the weak-interaction regime ($U<\Gamma_S$) where Andreev transport plays a dominant role, the local conductance $G_{LL}$ increases because the magnetic field shifts the internal quasiparticle peaks, $E^\uparrow_{A-}$ and $E^\downarrow_{A+}$, to their crossing. Next, for $b>b_{crit}$, the local conductance monotonously decreases because the quasiparticle states move aside to higher energies, reducing the zero-energy spectral function. During this changeover the nonlocal conductance $G_{RL}$ reverse its sign from negative to positive value, signaling the destructive influence of  magnetic field on the superconducting proximity effect. For stronger magnetic field, $b>b_{crit}$, the nonlocal conductance $G_{RL}$ monotonously decreases in much the same way as $G_{LL}$.

\begin{figure}[b!]
\centering
\includegraphics[width=0.95\linewidth]{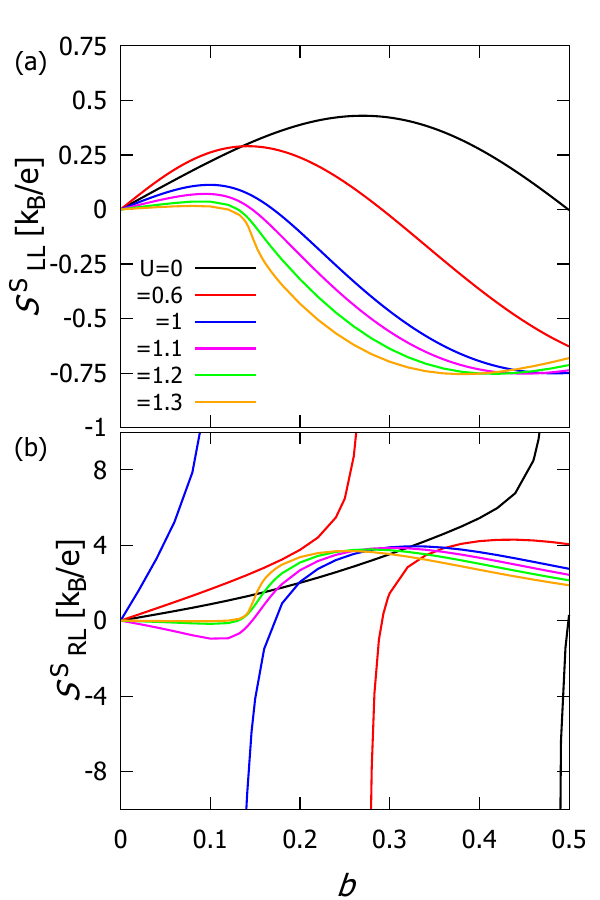}
\caption{
(a) The local $S^S_{LL}$  and (b) nonlocal $S^S_{RL}$ spin-dependent thermopower, varying against the Zeeman energy $b$ for selected values of the Coulomb interaction (as indicated) obtained for the same set of model parameters as in Fig.~\ref{figGLLRLkBB}.
}
\label{figSSLLSSRLkBB}
\end{figure}

The linear conductance of half-filled QD is spin-independent, $G^\uparrow_{\alpha\beta}=G^\downarrow_{\alpha\beta}$, therefore the spin polarization vanishes. In contrast, the spin-resolved Seebeck coefficients are antisymmetric, $S^\uparrow_{\alpha\beta}=-S^\downarrow_{\alpha\beta}$, implying $S_{\alpha\beta}=0$ and $S_{\alpha\beta}^{S}\neq 0$. Fig.~\ref{figSSLLSSRLkBB} displays variation of these spin-polarized coefficients $S_{LL}^{S}$ and $S_{RL}^{S}$ on $b$. For the Zeeman energy approaching the critical value, $b_{\mathrm{crit}}$, both $S_{LL}^{S}$ and $S_{RL}^{S}$ undergo the sign reversal. Furthermore, since the nonlocal thermopower depends on $G_{RL\sigma}$ (which vanishes when the opposite contributions from the ET and CAR channels compensate each other), the spin-resolved nonlocal thermopower $S_{RL\sigma}$ exhibits divergence in the weak-interaction regime. In the strong-interaction regime, this divergence is absent.

\section{Summary and outlook}
\label{sec:Conclusion}

We have investigated the local and nonlocal charge transport properties of a three-terminal hybrid junction, in which the correlated (Anderson-type) quantum impurity is side-coupled to a superconducting electrode and two metallic leads. Focusing on the linear-response regime, we have demonstrated that the singlet-doublet ground-state crossover leaves distinct signatures in both the local and nonlocal conductance, as well as in the thermopower. In particular, for values of the quantum-dot energy level and Coulomb interaction corresponding to the BCS-singlet ground state, the nonlocal conductance becomes negative. The transition from the doublet to the singlet regime is further signaled by a divergence of the nonlocal Seebeck coefficient and a sign reversal of the local thermopower. These effects originate from crossed electron-to-hole scattering processes, which dominate over the ballistic single-particle electron-transfer channel. 

We also predict characteristic features that emerge in the doublet--singlet crossover region as a consequence of the Kondo state induced by the exchange interaction between the quantum-dot spin and the spins of itinerant electrons in the normal leads \cite{Zitko-2015,Domanski-2016}. These features are manifested by an enhancement of the low-temperature conductance (Fig.~\ref{figGUTwnd1}) in the vicinity of the half-filling point ($\varepsilon \simeq -U/2$), as well as by a reversal of the sawtooth-like dependence of the local thermopower on the quantum-dot energy level $\varepsilon$ (Fig.~\ref{figSedprimkTS}). In contrast, the nonlocal transport properties turn out to be less sensitive to the subgap Kondo effect.

Our results, obtained using the self-consistent second-order perturbative treatment of the Coulomb interaction \cite{Yamada-2011}, are consistent with the local and nonlocal conductance calculated over a broad range of quantum-dot occupancies for the same nanostructure by means of unbiased numerical renormalization group (NRG) calculations in the $\Delta\rightarrow\infty$ limit \cite{Hashimoto-2024}. The present study extends these earlier results in several important respects: (i) it generalizes the analysis to the case of a finite superconducting gap $\Delta$, (ii) it provides complementary results for the Seebeck coefficients, and (iii) it highlights the important role of the subgap Kondo effect, as manifested in the temperature dependence of the transport properties, particularly the local ones. 

We also investigated the effect of the magnetic field on the quasiparticle spectrum and its impact on the transport properties. In the weakly interacting regime, the magnetic field primarily induces Zeeman splitting of the in-gap quasiparticle states, whereas in the strongly correlated regime, it suppresses the Kondo state and leads to a substantial rearrangement of the subgap quasiparticle spectrum. These effects are clearly reflected in all the transport characteristics.

We hope that our study will stimulate experimental investigations of the local and nonlocal conductance and thermopower as probes of the quantum-dot ground-state configuration. Such studies could be extended to more complex superconducting hybrid nanostructures containing two or more quantum dots, with the aim of exploring topologically nontrivial states of matter. Another challenging direction would be to investigate the transport properties of three-terminal superconducting nanostructures beyond the linear-response regime. Progress in this direction has recently been reported by A.~Oguri \textit{et al.}~\cite{Oguri-2026}, who predicted three-body correlations manifested in the quadratic dependence of the transport coefficients on energy, temperature, and bias voltage.

\begin{acknowledgements}
T.D. acknowledges support by the National Science Centre (Poland) within the Weave-Unisono grant no. 2022/04/Y/ST3/00061. V.P. acknowledges support from the Czech Science Foundation through Project No.~23-05263K and  European Cooperation in Science and Technology (EU COST) action CA21144 SUPERQUMAP.
\end{acknowledgements}

\section*{Data availability}
The datasets generated and analyzed during the current study are available from the repository \cite{data_2026}.

\appendix

\section{Influence of the pairing gap}
\label{gap_dependence}

\begin{figure}[t!]
\centering
\includegraphics[width=0.8\linewidth]{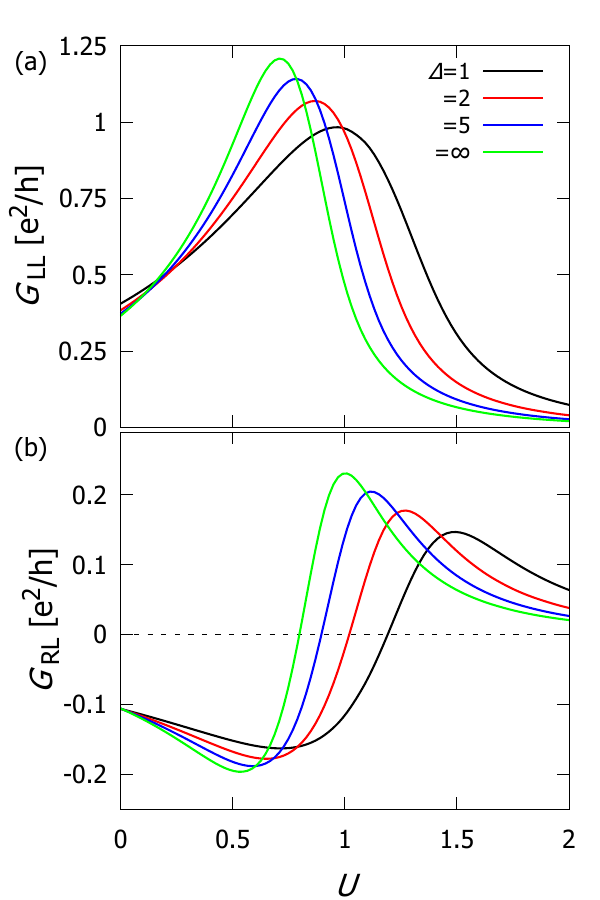}
\caption{
The (a) local $G_{LL}$  and (b) nonlocal $G_{RL}$ linear conductivity versus the Coulomb potential $U$ obtained for different pairing gap $\Delta$, using the same set of model parameters as in Fig.\ \ref{figGUTwnd1} for temperature $k_BT=0.05$.
}
\label{figGUdeltand1}
\end{figure}

\begin{figure}[b!]
\centering
\includegraphics[width=0.95\linewidth]{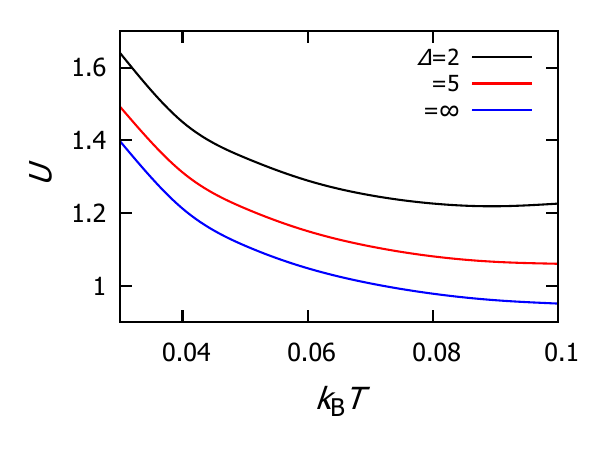}
\caption{
Temperature-dependent value of the Coulomb potential, $U$ [in units of $\Gamma_{S}\equiv 1$], at which  the local thermopower, $S_{LL}$, changes its sign. Results are obtained for $\varepsilon'=0.05$, $\Gamma_{NL}=\Gamma_{NR}=0.125$, and several values of the pairing gap $\Delta$, as indicated. 
}
\label{figUcredprim}
\end{figure}

Figure\ \ref{figGUdeltand1} displays the local and nonlocal conductivity as a function of the Coulomb interaction $U$ for different values of the pairing gap $\Delta$. We notice, that increase of $\Delta$ shifts the optimal values of both conductances toward the weaker Coulomb potential. In particular, for $\Delta \rightarrow \infty$ the maximal value of $G_{RL}$ tends to $U=\Gamma_{S}$, what coincides with the singlet-doublet transition in the superconducting atomic limit where the Kondo peak is strongly enhanced \cite{Domanski-2016}.

Figure \ref{figSedprimkU} (Sec.~\ref{sec:IIIB}) indicates sign change of the local thermopower $S_{LL}$ at some temperature-dependent value of the Coulomb potential $U$ in units of $\Gamma_S$. To address this phenomenon in more detail, we present in Fig.\ \ref{figUcredprim} variation of this Coulomb potential against temperature for slightly particle-hole asymmetric case, $\varepsilon'=0.05$, and several values of the pairing gap. The Coulomb potential of a given physical system is fixed, but the hybridization strength $\Gamma_{S}$ can be experimentally varied.

\begin{figure*}[t!]
\centering
\includegraphics[width=0.95\linewidth]{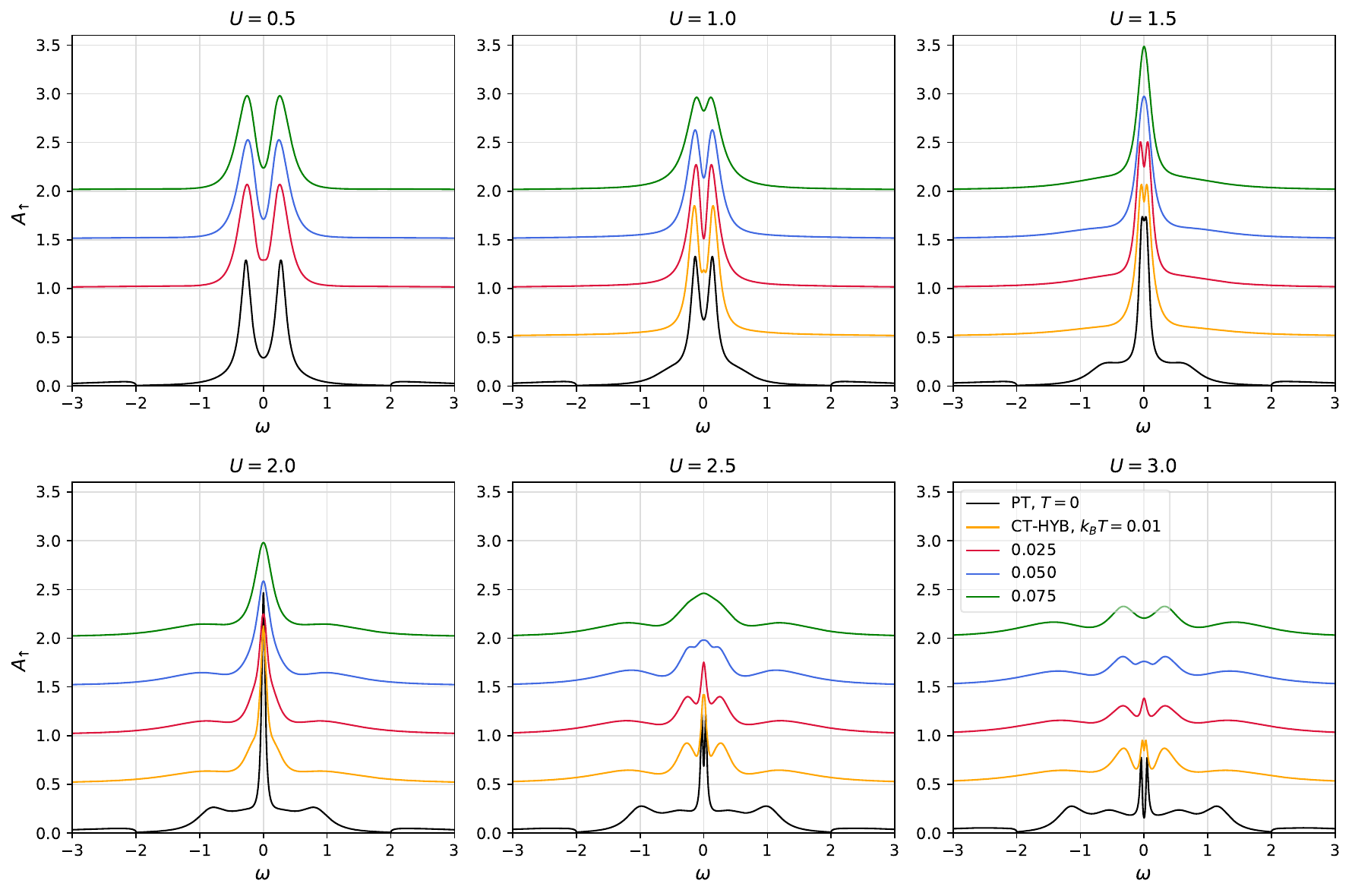}
\caption{
Spectral function of the half-filled quantum dot, $\varepsilon=-U/2$, obtained in absence of magnetic field, $b=0$, by maximum entropy method from the unbiased CT-HYB method results for several values of the Coulomb potential and temperatures, as indicated  [in units of $\Gamma_{S}$], compared to the perturbation theory (PT) result. The spectra corresponding to different temperatures are vertically shifted for clarity.}
\label{QMC}
\end{figure*}

\section{Unbiased QMC simulations}
\label{sec:QMC}

For the half-filled QD we have done the continuous-time hybridization expansion (CT-HYB) QMC simulation, using the TRIQS/CTHYB 3.2.0 solver \cite{Ref2}. As the Hamiltonian does not conserve charge due to the superconducting pairing term, we performed a canonical transformation in which electrons are mapped into holes in the spin-down segment~\cite{Pokorny-2018},
 $d_{\downarrow}^{\dagger} \rightarrow \tilde{d}_{\downarrow}$,
$c^{\dagger}_{\bk\downarrow S}\rightarrow \tilde{c}^{\phantom{\dagger}}_{\bk\downarrow S}$
and same for annihilation operators.
This transformation maps the superconducting pairing term 
$\Delta c^{\dagger}_{-\bk\uparrow S}c^{\dagger}_{\bk\downarrow S}$
to a transversal magnetic field 
$\Delta \tilde{c}^{\dagger}_{-\bk\uparrow S}\tilde{c}^{\phantom{\dagger}}_{\bk\downarrow S}$,
the local energy level $\varepsilon$ into an effective magnetic field and vice versa and it switches the sign of the interaction to attractive. The transformed Hamiltonian conserves the particle number and can be solved using standard Monte Carlo techniques.

As we solve the model for a half-filled orbital, it is convenient to transform the basis using unitary transformation \cite{Zalom-2021},
\begin{eqnarray}
\left( \begin{array}{c}
w_{\uparrow}\\
w_{\downarrow}
\end{array}\right) = 
\frac{1}{\sqrt{2}}
\left( \begin{array}{cc}
1 & \phantom{-}1\\
1 & -1
\end{array}\right)
\left( \begin{array}{c}
\tilde{d}_{\uparrow}\\
\tilde{d}_{\downarrow}
\end{array}\right) .
\end{eqnarray} 
This transformation mixes the off-diagonal parts of the Hamiltonian with the diagonal parts, and for half-filling it
leads to a diagonal model which can be solved more efficiently.

For computations we used the flat density of states of the external leads, with half-bandwidth $W/\Gamma_{S} = 100$. We used a 64 core computational node to run CT-HYB simulations. The results were obtained from $6.4 \times  10^7$ measurements with 200 Monte Carlo updates between measurements to minimize autocorrelation errors.

The spectral functions were obtained using analytic continuation from Matsubara to real frequencies $G_{11}(i\omega_{n}) \rightarrow G_{11}(\omega + i0^{+})$ using maximum entropy method \cite{Jarrell-1996} as implemented in the ana\_cont package \cite{Ref5}. We used a cutoff in Matsubara frequencies $\omega_{n,max} \leq 30$, constant default model of half-width $D/\Gamma_S = 10$ and the optimal value of the $\alpha$ hyperparameter was obtained using the {\it chi2kink} method \cite{Tremblay-2016}. 

\begin{figure*}[th]
\centering
\includegraphics[width=1\linewidth]{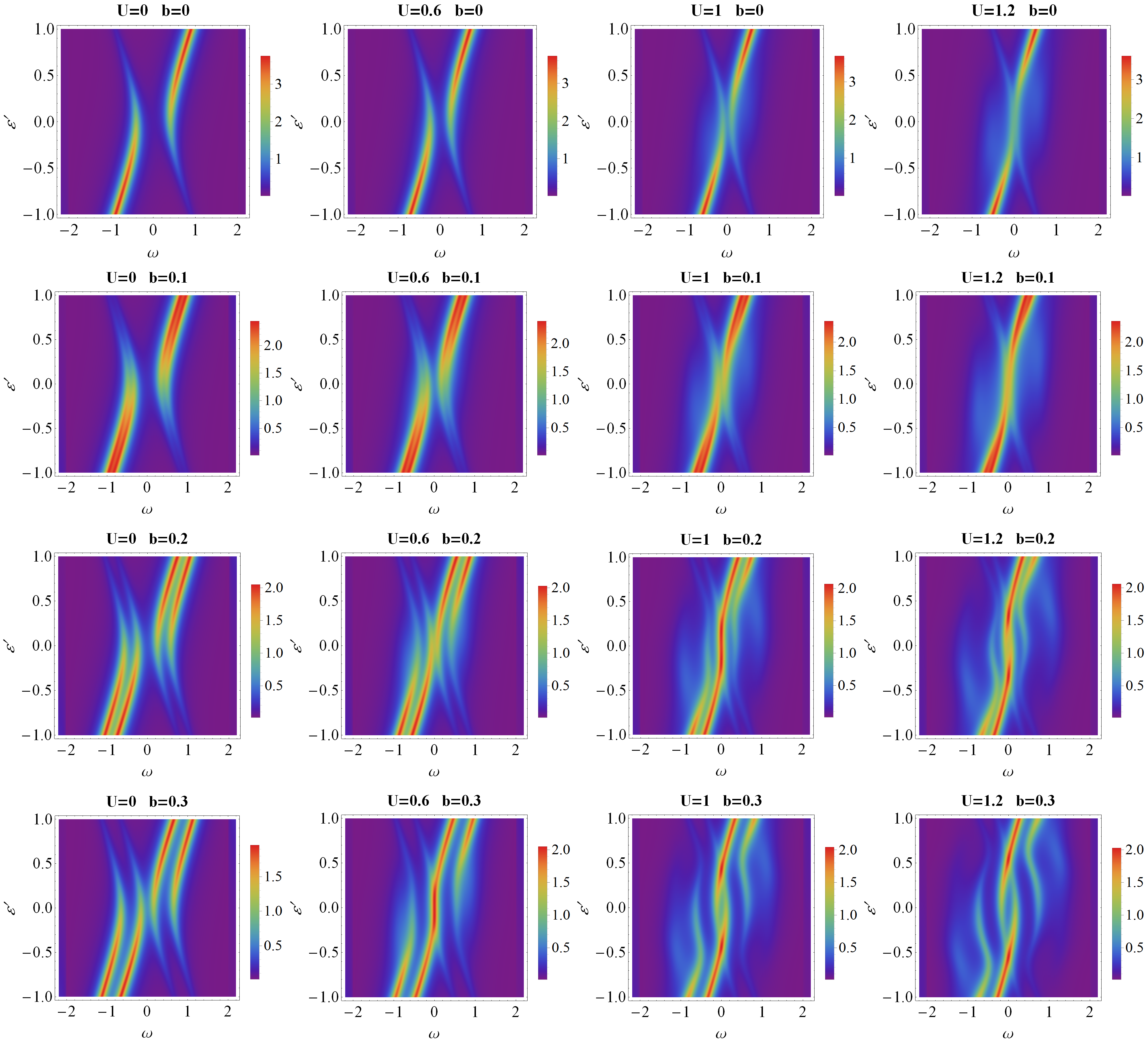}
\caption{
Normalized spectral function as a function of the detuning parameter $\varepsilon'$ obtained for several values of the Zeeman energy $b$ and Coulomb interaction $U$, using the model parameters $\Gamma_{S}=1$, $\Gamma_{NL}=\Gamma_{NR}=0.125$, $\Delta=2$ and $k_BT=0.05$.
}
\label{figAvedpkB}
\end{figure*}

Our results, displayed in Fig. \ref{QMC}, serve as a benchmark for the perturbative method which is used in the main part of this manuscript. We notice fairly satisfactory quantitative agreement between both approaches in the weak interaction regime, and quite good qualitative similarity even for the stronger interactions. The gap edge feature is smeared in QMC data as side-effect of the analytical continuation procedure. For the same reason there appear broad shoulders of the in-gap quasiparticles formed near $\omega=0.5$.

\section{Magnetic field effects for arbitrary filling of QD}
\label{Appendix_non_halfilled_QD}

\begin{figure*}
\centering
\includegraphics[width=1\linewidth]{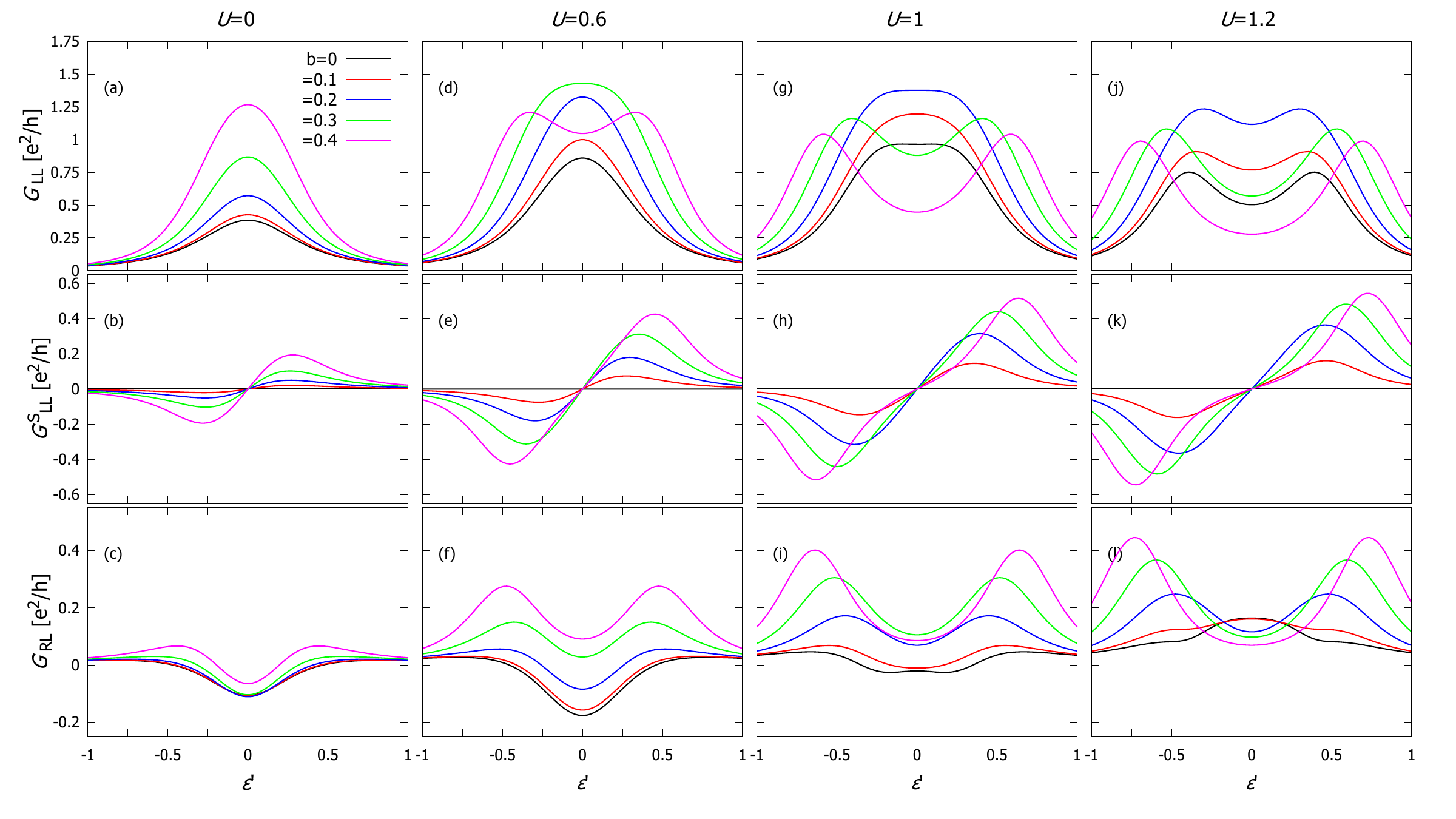}
\caption{
 The local linear conductance $G_{LL}$ (top row), polarization of local conductance $G^S_{LL}$ (middle row) and the nonlocal linear conductance $G_{RL}$ (bottom row) as a function of $\varepsilon'$ for the same set of parameters as in Fig.~\ref{figAvedpkB}. 
}
\label{figGLLRLked}
\end{figure*}

We now analyze the transport properties obtained numerically for an arbitrary quantum dot level position, $\varepsilon^{'}$, where $\frac{1}{2}\left(\varepsilon_{\uparrow}+\varepsilon_{\downarrow}\right)\equiv  \varepsilon^{'}-\frac{U}{2}$. Figure~\ref{figAvedpkB} shows the evolution of the quasiparticle spectrum as a function of $\varepsilon^{'}$ for representative values of the Coulomb interaction and magnetic field. In the weakly interacting regime, the magnetic field simply splits the Andreev bound states by the Zeeman energy. In contrast, in the strongly correlated regime, the magnetic field has a much more intricate effect. The bound states, which initially cross at $\varepsilon^{'}=0$ (see the panel for $U=1.2$, $b=0$), evolve into a four-peak structure when the magnetic field is applied (see the panel for $U=1.2$, $b=0.4$). Furthermore, the crossing point of the inner quasiparticle branches shifts to finite values of $\varepsilon^{'}$. This behavior indicates that the magnetic field modifies the singlet-doublet transition (crossover), as previously reported by Hashimoto \textit{et al}~\cite{Hashimoto-2024}.

Influence of the magnetic field on the local and nonlocal conductance as a function of the detuning parameter $\varepsilon'$ is presented in Fig.~\ref{figGLLRLked} for several values of the Zeeman energy and the Coulomb potential. The most significant changes are observed in the vicinity of the half-filling point, $\varepsilon'=0$. For weak magnetic fields, the local conductance is enhanced in entire region of $\varepsilon'$ (see the top-left panels of Fig.~\ref{figGLLRLked}). For stronger magnetic fields, $G_{LL}$ is suppressed near the region of the half-filled QD, as shown in Fig.~\ref{figGLLRLked}(j). Effectively, the local conductance reveals in the strong-correlation regime the two-peak structure, emerging at finite $\varepsilon'$.

In the presence of  magnetic field and away from the half-filling point, the spin-resolved local conductances become unequal, $G^\uparrow_{LL}\neq G^\downarrow_{LL}$, giving rise to a finite spin conductance $G^S_{LL}$ (see the middle row of Fig.~\ref{figGLLRLked}). Since the direct and crossed Andreev reflection coefficients, $T^\mathrm{DAR}_{LL\sigma}$ and $T^\mathrm{CAR}_{LR\sigma}$, are spin independent, the sign of $G^S_{LL}$ is determined solely by a contribution of the electron-transfer (ET) channel to $G^\sigma_{LL}$. Consequently, $G^S_{LL}=e\left(D_{LR,\mu_\up}^\tx{ET}-D_{LR,\mu_\down}^\tx{ET}\right)$.  
For $\varepsilon'<0$, the spin-resolved spectral function  $A_{\downarrow}(\omega=0)>A_{\uparrow}(\omega=0)$, which implies $G^\downarrow_{LL}>G^\uparrow_{LL}$ and, consequently, a negative spin conductance, $G^S_{LL}<0$. On the other hand, for $\varepsilon'>0$, the magnitudes of spin-resolved spectral functions are reversed, resulting in a positive spin conductance, $G^S_{LL}>0$.

Magnetic field modifies the coefficient $D_{LR,\mu_\sigma}^\tx{ET}$ so, consequently, the nonlocal conductance $G_{RL}$ becomes sensitive to the Zeeman energy (see the bottom row of Fig.~\ref{figGLLRLked}). In the weak interaction regime, the application of the magnetic field induces a sign reversal of $G_{RL}$, indicating that the ET contribution to the conductance eventually exceeds that of the CAR channel. In contrast, for the strong Coulomb potential, the nonlocal conductance remains positive over the entire range of $\varepsilon'$ and $b$. In this regime, the strong magnetic field suppresses $G_{RL}$ in the vicinity of $\varepsilon'\approx 0$, whereas for large values of $\left|\varepsilon'\right|$ the nonlocal conductance is enhanced.

\begin{figure*}
\centering
\includegraphics[width=1\linewidth]{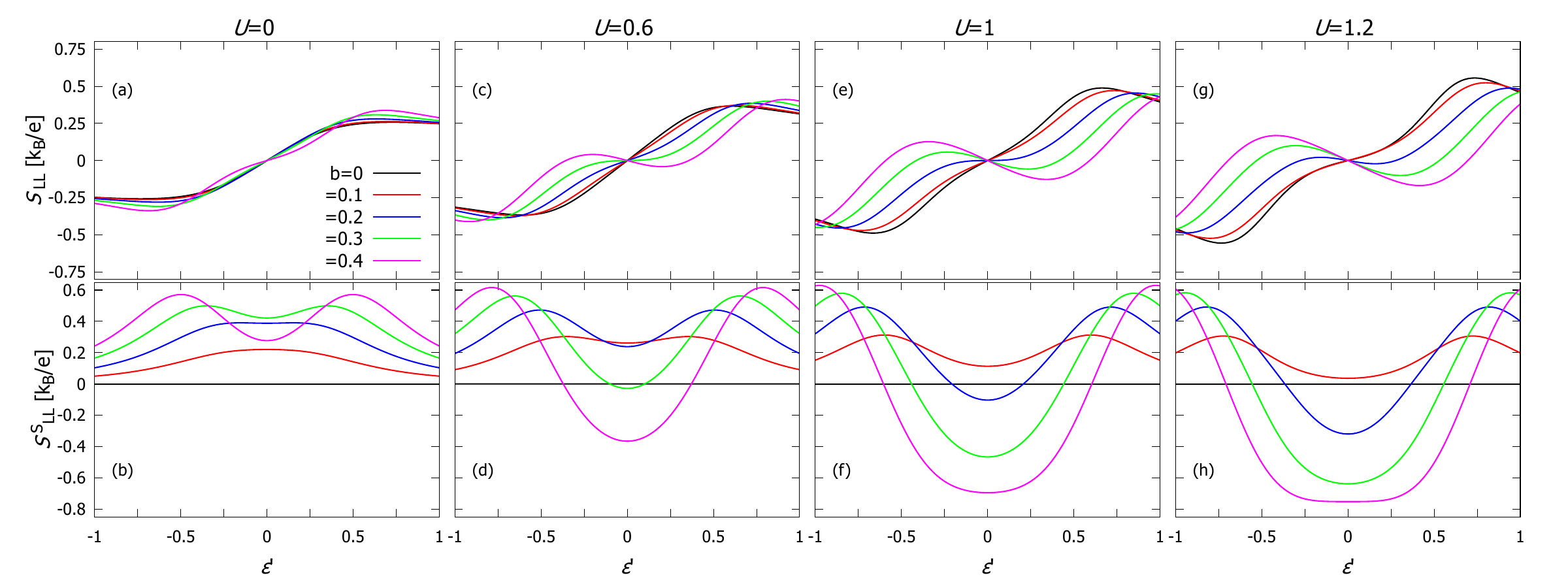}
\caption{
 The local thermopower $S_{LL}$ (top row) and the spin thermopower $S^S_{LL}$ (bottom row) as functions of the detuning parameter $\varepsilon'$ for several values of the Zeeman energy $b$ and Coulomb interaction $U$, assuming $\Gamma_{S}=1$, $\Gamma_{NL}=\Gamma_{NR}=0.125$, $\Delta=2$, and $k_BT=0.05$.
}
\label{figSLLSSLLked}
\end{figure*}

The top row of Fig.~\ref{figSLLSSLLked} shows the local thermopower $S_{LL}$ as a function of the detuning parameter $\varepsilon'$ for several values of the Coulomb interaction and Zeeman energy. The sign of $S_{LL}$ indicates which type of charge carrier (electrons or holes) dominates the transport. A negative value of $S_{LL}$ corresponds to electron-dominated transport, whereas a positive value indicates that holes play the dominant role. For weak magnetic fields, we obtain $S_{LL}<0$ for $\varepsilon'<0$ and $S_{LL}>0$ for $\varepsilon'>0$. However, when the Zeeman energy exceeds a critical value, $b>b_{\mathrm{crit}}$, the thermopower undergoes a sign reversal.

Magnetic field strongly affects also the spin thermopower, $S^S_{LL}$, as shown in the second row of Fig.~\ref{figSLLSSLLked}. For weak magnetic fields, $S^S_{LL}$ is positive, whereas for $b>b_{\mathrm{crit}}$ it becomes negative in the vicinity of the half-filling point.

\bibliography{myBib}

\end{document}